\def\extended{true}
\def\extendedyes{true}
\ifx\extended\extendedyes
  \documentclass[sigconf,nonacm]{acmart}
\else
  \documentclass[sigconf]{acmart}
\fi

\usepackage{paralist}
\usepackage{algorithm}
\usepackage{algpseudocode}
\usepackage{dsfont}
\usepackage{subcaption}
\usepackage{enumitem}
\usepackage{hyperref}
\usepackage{multirow}
\usepackage{graphicx}
\usepackage{booktabs}
\usepackage{pifont}
\usepackage{ifthen}
\usepackage[table]{xcolor}

\renewcommand\leq\leqslant
\renewcommand\geq\geqslant
\newcommand{\cls}[1]{\textsc{\small #1}}

\newcommand{\datasetname}{\textsc{StatSheets}}
\newtheorem{definition}{Definition}
\newtheorem{problem}[definition]{Problem}

\newcommand{\bparagraph}[1]{\vspace{1.5mm}\noindent{\textbf{\upshape
#1}}\par\noindent\ignorespaces}

\newcommand{\printIfExtVersion}[2]{\ifthenelse{\equal{\extended}{true}}{#1}{#2}}
\newcommand{\versioning}[2]{\printIfExtVersion{#2}{#1}}
\newcommand{\cmark}{\ding{51}}
\newcommand{\xmark}{\ding{55}}

\algtext*{EndWhile}
\algtext*{EndFor}
\algtext*{EndIf}

\printIfExtVersion{}{%
\copyrightyear{2026}
\acmYear{2026}
\setcopyright{cc}
\setcctype{by}
\acmConference[CIKM '26]{Proceedings of the 35th ACM International Conference on Information and Knowledge Management}{November 07--11, 2026}{Rome, Italy}
\acmBooktitle{Proceedings of the 35th ACM International Conference on Information and Knowledge Management (CIKM '26), November 07--11, 2026, Rome, Italy}
\acmDOI{10.1145/3799682.3840907}
\acmISBN{979-8-4007-2539-5/2026/11}
}

\begin{document}

\ifthenelse{\equal{\extended}{true}}{
\title{Structured Prediction for Scalable Spreadsheet Table Understanding: From Cell Types to Table Ranges\\(Extended Version)}
}{
\title{Structured Prediction for Scalable Spreadsheet Table Understanding: From Cell Types to Table Ranges}
}

\author{Antoine Gauquier}
\email{antoine.gauquier@ens.psl.eu}
\orcid{0009-0005-9573-6364}
\affiliation{%
        \institution{DI ENS, ENS, CNRS, PSL, Inria}
        \city{Paris}
        \country{France}
}

\author{Ioana Manolescu}
\email{ioana.manolescu@inria.fr}
\orcid{0000-0002-0425-2462}
\affiliation{%
  \institution{\makebox[0pt]{Inria \& Institut Polytechnique de Paris}}
        \city{Palaiseau}
        \country{France}
}

\author{Pierre Senellart}
\email{pierre@senellart.com}
\orcid{0000-0002-7909-5369}
\affiliation{%
        \institution{DI ENS, ENS, CNRS, PSL, Inria}
        \city{Paris}
        \country{France}
}

\renewcommand{\shortauthors}{Antoine Gauquier, Ioana Manolescu, and Pierre Senellart}

\begin{abstract}
  Spreadsheets are a primary medium for publishing tabular data, yet automatically extracting structured content from them remains
  difficult due to heterogeneous layouts, diverse file formats, and inconsistent organizational conventions. We address two core
  tasks in spreadsheet understanding: \emph{Cell-Type Classification} (CTC), which assigns roles to cells, and \emph{Table Detection}
  (TD), which identifies table bounding boxes within sheets.
  We propose an efficient two-stage pipeline in which a learned CTC model feeds a deterministic TD algorithm. For CTC, we use a LightGBM
  classifier over 65 structured features together with a pairwise CRF enforcing spatial consistency across the cell grid. Our TD method
  extracts table ranges from predicted cell types by a deterministic five-stage procedure.
  For evaluation, we built and share \datasetname{}, a multilingual benchmark of 737 manually annotated sheets from 14 public data providers
  across multiple countries and file formats. Under 5-fold cross-validation, our \cls{CRF-LightGBM} system achieves a Mean File-Macro
  F$_1$ score of 0.937 on CTC, within 0.6 percentage points of the GPU-based \cls{TUTA} Transformer, while requiring substantially fewer
  computational resources. For TD, our deterministic approach outperforms region-based baselines and remains competitive with recent
  LLM-based systems such as \cls{SpreadsheetLLM}. These results demonstrate that combining non-linear structured prediction with
  deterministic range extraction provides a competitive, scalable, and computationally efficient approach to spreadsheet table understanding.
\end{abstract}



\maketitle

\setlength{\abovedisplayskip}{4pt plus 0pt minus 2pt}
\setlength{\belowdisplayskip}{4pt plus 0pt minus 2pt}
\setlength{\abovedisplayshortskip}{4pt plus 0pt minus 2pt}
\setlength{\belowdisplayshortskip}{4pt plus 0pt minus 2pt}

\section{Introduction}
\label{sec:introduction}

Spreadsheets are among the most widely used data management artifacts. Governments, international organizations,
companies, and journalists routinely publish data in formats such as XLSX, XLS, ODS, CSV, and TSV.
Public repositories such as \href{https://ec.europa.eu/eurostat}{Eurostat} alone expose thousands of spreadsheets, many continuously updated
and reused in downstream pipelines~\cite{gauquier2026efficientandscalable}. Unlike relational databases, however,
\emph{spreadsheets are designed for human interpretation rather than machine consumption}: their semantic structure is
only partially explicit and instead conveyed through layout, formatting, merged cells, spatial proximity, etc. The conventions for doing so
vary across publishers, domains, languages, and software ecosystems.

This flexibility, while central to spreadsheets' success, makes automated extraction and integration difficult. Real-world spreadsheets rarely
contain clean rectangular tables. Instead, they mix data regions with hierarchical headers, titles, footnotes, metadata, and other auxiliary
elements within the same sheet. %
Similar layouts may encode different semantics, while equivalent structures may appear in very different forms. Structural irregularities such as sparsity, merged cells, and discontinuous regions further complicate the problem.
Consequently, recovering machine-interpretable tables from spreadsheets is substantially harder than extracting structured data from HTML tables or database exports.

Robust, large-scale spreadsheet understanding is increasingly critical for modern data management systems, including data lake construction, open-data indexing, fact-checking,
business intelligence, and retrieval-augmented analytics. Errors in the detection of table boundaries or cell roles propagate to downstream tasks,
producing corrupted schemas, invalid joins, or unusable datasets: spreadsheet understanding is not merely a preprocessing step, but \emph{a foundational
problem for scalable structured-data acquisition}.
Two core tasks can be identified: \emph{Cell-Type Classification} (CTC) assigns semantic roles (e.g., \cls{HEADER}, \cls{DATA}, \cls{TITLE}) to cells,
while \emph{Table Detection} (TD) recovers table boundaries within sheets. Although CTC and TD  have received continuous attention during the last decade, see~\cite{chen2014integrating,koci2016machine,gonsior2020active,ghasemi2019tabular,dong2019semantic,sun2021hybrid,wang2021tuta},
and~\cite{coletta2012public,vitagliano2021detecting,koci2018table,koci2019genetic,dong2019tablesense,dong2024encoding}, respectively, existing approaches still face important limitations.
\begin{inparaenum}[\bfseries\arabic*.] \item Recent progress relies on large neural architectures \emph{trained on proprietary corpora} and which are \emph{expensive}. Transformer-based systems such as \cls{TUTA}~\cite{wang2021tuta}
and LLM-based approaches such as \cls{SpreadsheetLLM}~\cite{dong2024encoding} achieve strong results through large-scale pre-training, %
but at substantial %
cost (Sec.~\ref{sec:exp_ctc} and~\ref{sec:exp_td}). Such GPU-intensive systems
can be problematic for %
pipelines processing thousands to millions of spreadsheets, where throughput and
operational cost remain primary constraints.
\item Prior work \emph{only partially captures the heterogeneous signals} governing spreadsheet structure. Some methods emphasize semantic embeddings while under-exploiting
formatting and geometric regularities; others focus on visual or region-based segmentation while insufficiently modeling long-range dependencies. Yet, {spreadsheet semantics
emerges precisely from the interaction of multiple weak signals}, including lexical content, formatting consistency, row- and column-level regularities, neighborhood coherence, and
global spatial organization.
\item \emph{Reproducibility and evaluation remain problematic}. Several influential systems do not release implementations~\cite{dong2019semantic, sun2021hybrid, koci2018table, koci2019genetic, dong2019tablesense}, or
rely on proprietary datasets~\cite{dong2019semantic, dong2019tablesense, dong2024encoding}, preventing direct comparison and independent validation. Thus, it remains difficult to isolate
the respective contributions of semantic modeling, structural priors, feature engineering, and computational scale.
These issues are compounded by the lack of public datasets jointly supporting CTC and TD. Existing benchmarks are either single-task, structurally limited, or derived from outdated corpora.
\cls{DECO}~\cite{koci2019deco}, the only public benchmark covering both tasks, is based on the Enron archive~\cite{hermans2015enron} and reflects practices from the early 2000s;
it excludes large spreadsheets,
lacks multilingual coverage, and omits visually hidden content, making it poorly suited for evaluating modern structured-prediction methods.
\end{inparaenum}

In this work, we make the following contributions:

\noindent (1)\, We introduce \datasetname{}, a \textbf{publicly available dataset of 737 manually annotated spreadsheet sheets} collected from 14 public organizations publishing statistical data
across multiple countries, languages, and spreadsheet formats. Unlike prior datasets, \datasetname{} jointly supports both CTC and TD under realistic conditions,
including large sheets, multilingual content, heterogeneous layouts, and multiple spreadsheet formats (Sec.~\ref{sec:dataset}).

\noindent (2)\, We propose an \textbf{efficient end-to-end spreadsheet understanding pipeline combining structured cell-type prediction with deterministic table extraction}
(Sec.~\ref{sec:approach}), which operates in two sequential stages. First, a non-linear classifier predicts per-cell roles from a rich feature space
combining lexical, formatting, positional, and row-/column-aware structural descriptors; these predictions are then
refined through a pairwise Conditional Random Field (CRF) enforcing spatial coherence over the spreadsheet grid (Sec.~\ref{sec:ctc_pipeline}). Second, table
ranges are recovered through a deterministic 5-stage extraction algorithm operating on predicted cell-type grids (Sec.~\ref{sec:approach_td}).

\noindent (3)\, We conduct a \textbf{thorough experimental study} of both tasks, comparing against non-DL and DL state-of-the-art approaches (Sec.~\ref{sec:experiments}).
Our experiments show that:
  \begin{inparaenum}[(i)]
  \item on CTC, our \cls{CRF-LightGBM} pipeline achieves a Mean File-Macro F$_1$ of 0.937, within 0.6 pp of the Transformer-based \cls{TUTA} model, while
  requiring substantially lower computational resources and avoiding GPU dependence (Sec.~\ref{sec:exp_ctc});
  \item on TD, our deterministic extraction algorithm consistently outperforms generic region-based methods and remains competitive with recent LLM-based
  systems, again with orders of magnitude of computational cost savings (Sec.~\ref{sec:exp_td}).
  \end{inparaenum}
  These results show that \textbf{carefully designed structured models, combining non-linear feature modeling with structured spatial inference, remain highly competitive
for spreadsheet understanding}, particularly in realistic large-scale data-management settings \textbf{where scalability, robustness, and interpretability are critical}.

We discuss related work in Sec.~\ref{sec:related_work}. The \datasetname{} dataset, our implementation of \cls{SpreadsheetLLM}, and the
code to reproduce the experiments are available~at~\cite{github_repository}. 
\versioning{An extended version of this paper is available in~\cite{gauquier2026structuredextended}.}{This paper is an extended version of the conference paper~\cite{gauquier2026structured}.}
\section{Related Work}
\label{sec:related_work}

\bparagraph{Cell-Type Classification (CTC) in Spreadsheets} Methods
in this area have evolved from feature-engineered, to large pre-trained neural
architectures. Early work
\cite{chen2014integrating} uses an undirected graphical model combining formatting and content cues to recover
annotation-to-data mappings for relational integration, with semi-automatic error repair, emphasizing spatial dependencies
but requiring human interaction and targeting relational extraction rather than general-purpose CTC.
\cite{koci2016machine} introduces the first dedicated CTC framework using a Random Forest (RF) over formatting,
content, and positional features to classify cells independently, later extended with active learning by \cite{gonsior2020active}.
Our \cls{RF-Koci} baseline follows~\cite{koci2016machine} with a similar feature design.
\cite{christodoulakis2020pytheas}~introduces \cls{Pytheas}, a fuzzy-rule-based system classifying CSV rows
into roles such as header, data, and context from column coherency and semantic keywords.
Its formulation assigns a single label to
each spreadsheet row. Thus, it cannot represent multiple
tables appearing side by side, making precise 2-dimensional TD impossible.
It also precludes cell-level CTC, since all cells in
a row share the same label, despite spreadsheets commonly mixing roles
within rows (e.g., header then data cells).
Recent works use DL approaches with semantic and contextual representations. \cite{ghasemi2019tabular} combines
pre-trained cell embeddings, recurrent architectures and formatting features. \cite{dong2019semantic} proposes a multi-task
framework jointly addressing TD, structural component recognition, and
CTC using BERT embeddings,
while \cite{sun2021hybrid} combines neural embeddings with Probabilistic Soft Logic constraints
encoding structural regularities. Neither released code nor datasets.
\cls{TUTA}~\cite{wang2021tuta} extends Transformers with structure-aware self-attention capturing spreadsheet
layout and semantics, pre-trained on large spreadsheet and Web table corpora. It substantially outperforms earlier neural methods;
we therefore use it as our main DL baseline (Sec.~\ref{sec:exp_ctc}).
\cls{ForTaP}~\cite{cheng2022fortap} extends \cls{TUTA} with formula-aware pre-training
objectives, but relies on %
expert-authored formulas.

\bparagraph{Table Detection (TD) in Spreadsheets} We identify two categories: \emph{region-based} methods (extracting generic
 grid structures) and \emph{table-specific} ones (modeling table semantics and structure
to extract table boundaries).

\emph{Region-based approaches.}
\cite{coletta2012public} groups adjacent non-empty cells into connected components represented by bounding
boxes. A purely structural, learning-free approach targeting %
dense regions rather than tables, it is sensitive to sparsity and cannot distinguish tables from %
metadata. \cite{vitagliano2021detecting} introduces \cls{Mondrian}, which represents spreadsheets as binary images
segmented into homogeneous rectangular regions through density-based clustering and compares regions across
files %
to identify recurring templates. Although learning-free and unsupervised,
both %
methods detect generic structural regions rather than tables.
Their evaluation in \emph{region-anchored mode} overestimates true TD quality;
as shown in Sec.~\ref{sec:exp_td}, performance drops sharply at higher IoU thresholds, highlighting
the limits of generic region extraction for precise TD.

\emph{Table-specific approaches.}
Other works explicitly focus on predicting table boundaries. \cite{koci2018table} proposes the graph-based \emph{Remove and Conquer} (RAC) framework,
which infers cell layout roles before constructing graphs over spatially related regions to detect table boundaries through curated rules. As one of
the earliest %
coupled CTC+TD pipelines, it is also the prior approach closest to ours.
\cite{koci2019genetic} extends RAC with a genetic optimization procedure %
maximizing a fitness function over predicted
cell-type grids. Although effective with accurate CTC predictions, it is sensitive to classification errors and noise%
~\cite{vitagliano2021detecting}.
Neither~\cite{koci2018table} nor \cite{koci2019genetic} releases code, datasets, or trained models, preventing direct comparison.
\cite{dong2019tablesense} introduces \cls{TableSense}, a CNN-based end-to-end TD framework formulating detection as %
object detection over featurized
cell grids. Trained on the proprietary \cls{WebSheet10K} dataset, %
it is reported as a strong specialized TD system, though no code, models, or datasets were released;
later, \cls{SpreadsheetLLM}, from the same Microsoft research group, reports substantially outperforming
it.
\cls{SpreadsheetLLM} \cite{dong2024encoding} encodes spreadsheets into
compact token sequences, %
and fine-tunes LLMs for TD and related tasks (details in
Sec.~\ref{sec:exp_td}). We include it as a direct competitor using our
own implementation, derived from the supplementary materials
of~\cite{dong2024encoding}.

\bparagraph{End-to-end Spreadsheet Understanding}
Few works address spreadsheet understanding as unified CTC+TD pipelines. \cite{koci2018table}, later extended by~\cite{koci2019genetic},
couples the tasks through rule-based extraction but releases neither code nor trained models.
\cite{koci2019xlindy} demonstrates \cls{XLIndy}, an interactive Excel add-in for layout inference and table recognition,
without an evaluation or sharing the code.
\cite{dong2019semantic} jointly addresses TD, structural component recognition, and CTC via a multi-task FCNN with
strictly coarse-to-fine dependencies: TD feeds component recognition, which then feeds CTC, without reverse interaction.
In contrast, our pipeline performs CTC based solely on the sheet %
and derives table boundaries deterministically and exclusively from the predicted cell-type grid. Moreover,
\cite{dong2019semantic} relies on a proprietary dataset and releases no implementation.
Overall, prior end-to-end systems rely on rule-based extraction, proprietary implementations, or coarse-to-fine designs
where TD is not influenced by downstream CTC. In contrast, we derive table boundaries solely from
CTC predictions while jointly evaluating both tasks under cross-validation.

\bparagraph{Datasets for Spreadsheet Understanding}
The only publicly available dataset covering both CTC and TD is \cls{DECO}~\cite{koci2019deco}, built from the Enron
email archive~\cite{hermans2015enron}. %
\cls{DECO} is unsuitable in
our setting for several reasons. First, it excludes files larger than
5~MB, while large tables are common in public
data portals, e.g., Eurostat tables regularly exceed hundreds
of MB~\cite{gauquier2026efficientandscalable}. Second, it excludes
non-Western languages such as Arabic and Japanese; this introduces semantic and structural biases (e.g., right-to-left layout in Arabic spreadsheets).
Third, based on Enron emails (2000--2001), it reflects outdated formatting practices. %
Finally, annotations only cover visible content: hidden rows/columns (via zero height/width or explicit hide/unhide)
are unlabeled, leaving large parts of sheets unannotated and making the dataset incompatible with spatial and sequential methods
that require complete grid annotations.
Several smaller CTC-only benchmarks also exist. \cls{DeEx}~\cite{eberius2013deexcelerator} provides 444
sheets with a six-type label scheme, while \cls{SAUS}~\cite{chen2014integrating} and \cls{CIUS}~\cite{ghasemi2019tabular}
follow the same scheme on narrow domains. However, all three provide only CTC labels, lack TD annotations,
and have limited linguistic, formatting, and structural diversity; \cls{DeEx} and \cls{SAUS} are no longer accessible.
The proprietary \cls{WebSheet10K} dataset used in~\cite{dong2019tablesense, wang2021tuta, dong2024encoding} %
is also unavailable publicly. A subset of related annotations (VEnron2, VEUSES, VFUSE) released by \cite{dong2019tablesense} contains only
TD table ranges (no CTC labels) and is restricted to English XLS/XLSX files derived from Enron/EUSES.
To facilitate extensive evaluation at scale, we therefore introduce \datasetname{} (Sec.~\ref{sec:dataset}), a publicly available dataset addressing these limitations:
737 sheets in six  languages across five file formats, with complete cell-level CTC and table-range TD annotations
collected from contemporary statistical spreadsheets published by international organizations.

\section{Our Approach}
\label{sec:approach}

\begin{figure*}
    \centering
    \includegraphics[width=\linewidth]{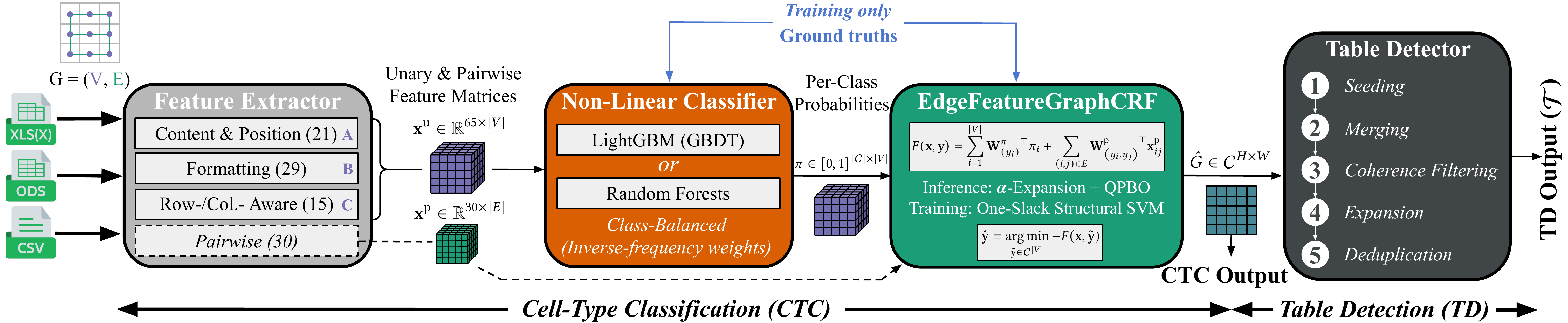}
    \vspace*{-2.5em}
    \caption{End-to-end \emph{Cell-Type Classification} (CTC) and \emph{Table Detection} (TD) pipeline.}
    \label{fig:pipeline}
\end{figure*}

We decompose spreadsheet understanding into two stages 
(Figure~\ref{fig:pipeline}). CTC (Sec.~\ref{sec:approach_ctc}) assigns each cell a label
using a non-linear %
classifier combined with a pairwise CRF enforcing
spatial coherence. TD (Sec.~\ref{sec:approach_td}) takes
the resulting label grid and recovers axis-aligned table extents through a
deterministic, learning-free algorithm.

\subsection{Cell-Type Classification}\label{sec:approach_ctc}

Let $\mathcal{C} = \{\cls{EMPTY}, \cls{HEADER},$ $\cls{DATA}, \cls{TITLE}, \cls{OTHER}\}$ be the set of five cell-type classes.
We formulate CTC as a structured prediction problem on a 2-dimensional grid,
where each cell is a node, and edges connect horizontal and vertical neighbors.
Each label $\hat{y}_i \in \mathcal{C}$ should be assigned considering
both local features and neighborhood consistency.
This motivates a two-stage pipeline in which a non-linear classifier produces
per-cell class probabilities that serve as unary potentials for a pairwise
CRF which enforces spatial coherence.

\begin{problem}[Cell-Type Classification]\label{problem:1}
   A spreadsheet region is represented as a
    2-dimensional grid $G = (V, E)$, where each node $i \in V$ corresponds to a cell and
    $E \subseteq V \times V$ is the set of pairs of horizontally or vertically adjacent cells.
    Given %
    $G$ and some features describing its elements, %
    the CTC problem consists of predicting a class for each cell,
    or, equivalently, predicting for all cells the label vector~$\hat{\mathbf{y}} \in \mathcal{C}^{|V|}$.
    \end{problem}

\bparagraph{Features}
\versioning{First, we rely on \emph{65 unary features} (65-dimensional vector) for
each cell, organized into three groups
(see Table~\ref{tab:ctc_unary_features}; we detail them in the
extended version~\cite{gauquier2026structuredextended}). \emph{Group A} contains content and positional features
describing intrinsic cell properties and local context, \emph{Group B} contains formatting and layout
features extracted from rich spreadsheet formats (XLSX, XLS, ODS), with default
values for plain-text formats (CSV, TSV). \emph{Group C} contains row-/column-aware
statistical features, capturing global structural~context.}{Each cell is represented by a 65-dimensional  unary feature vector, organized into three groups.
The design follows~\cite{koci2016machine}, and extends it with additional (richer) formatting attributes
and row-/column-level statistical features. Table~\ref{tab:ctc_unary_features} provides a summary.}

\printIfExtVersion{
\emph{Group A: Content and positional features (21 features, IDs 1--21).}
These features capture intrinsic cell properties and simple local context.
\emph{Type indicators} encode whether a cell is empty, numeric (integer/float), string, date, or formula.
\emph{Text statistics} describe the string representation: length, counts of digits/letters/spaces/other characters, and
binary start-with-letter/start-with-digit flags. %
\emph{Positional features} include normalized row/column indices within the non-empty region and
a boundary flag for sheet borders. %
\emph{Context features} capture local structure via the number of
non-empty 2-dimensional
neighbors, an isolation flag (surrounded by empty cells), and two flags distinguishing ODS from XLS/XLSX formats.

\emph{Group B: Formatting features (29 features, IDs 22--50).}
This group encodes visual and layout attributes from rich spreadsheet formats
(XLSX, XLS, ODS). Features include font weight, italic, underline, font
size, hashed font name, text/background colors
(RGB, normalized to $[0,1]^3$), a merged-cell flag, and four border flags
(left, right, top, bottom).
Horizontal and vertical cell alignments are one-hot encoded
(7 horizontal modes: general/start, right, center, justify, left, fill, distributed;
6 vertical modes: top, center, bottom, distributed/justify, baseline, other),
harmonized across file formats.
Extraction is implemented separately for XLSX (using \href{https://openpyxl.readthedocs.io/}{\emph{OpenPyXL}}), XLS (\href{https://xlrd.readthedocs.io/}{\emph{xlrd}}),
and ODS (\href{https://pypi.org/project/odfpy/}{\emph{ODFPy}}), but mapped to a shared
29-dimensional representation. Robust fallbacks handle missing or inconsistent style
metadata, common in real-world spreadsheets. For plain-text formats (CSV, TSV),
all formatting features are set to default values. %

\emph{Group C: Row- and column-aware features (15 features, IDs 51--65).}
These features capture global row/column-level structure beyond the local 4-neighborhood.
For each row, we compute
the fraction of non-empty, numeric, and string cells, a type-homogeneity
score (majority-type fraction), a type-uniformity flag (whether all non-empty
cells share the same type), and an isolation flag (only non-empty cell in the row). Analogous statistics are computed for columns, except for
type uniformity.
In addition, a column-type outlier flag indicates whether the cell's type differs from the column majority,
and 3 merge-span statistics capture the fractions of merged cells in the row and in the column,
and whether the cell's merged region spans at least three columns.
}{}

\emph{Pairwise features (30 features).}
For each edge $(i,j)$ in the 2D grid, we compute a %
feature vector describing the relationships between adjacent cells.
These features fall in four groups.
\emph{Orientation features} encode the relative horizontal and vertical positioning between cells.
\emph{Type features} encode agreement and transitions, including
whether cells share
the same type (empty, numbers, strings, dates, or formula) and type
transitions (empty--non-empty, number--string, number--date).
\emph{Content difference features} capture absolute differences in text
statistics (character and digit counts, jointly starting
with a letter, etc.).
\emph{Formatting features} measure visual agreement: same
font size and name, consistent bold/italic, same font and background
colors,
shared-border presence, horizontal/vertical alignment
differences, and whether both cells are part of merged~regions.

\begin{table*}
  \centering
  \caption{All 65 unary features used in the CTC task, organized into the three groups A, B, and C.}
  \vspace*{-1em}
  \label{tab:ctc_unary_features}
  \scriptsize
  \setlength{\tabcolsep}{4pt}
  \hspace*{-1.25em}
  \begin{tabular}{crlp{2.9cm}rlp{2.9cm}rlp{2.9cm}}
  \toprule
   & ID & Name & Description & ID & Name & Description & ID & Name & Description \\
  \midrule

  \multirow{7}{*}{\rotatebox{90}{\textbf{Group A}}}
  & 1 & \textsc{IsEmpty}    & No cell content
  & 8 & \textsc{\#Chars}      & Character count
  & 15 & \textsc{RowPosRatio} & Relative row index in grid \\

  & 2 & \textsc{IsNum}      & Value is numeric
  & 9 & \textsc{\#Digits}     & Digit count
  & 16 & \textsc{ColPosRatio} & Relative column index in grid \\

  & 3 & \textsc{IsStr}      & Value is a string
  & 10 & \textsc{\#Alpha}    & Alphabetic character count
  & 17 & \textsc{IsSheetBorder} & Cell on non-empty boundary \\

  & 4 & \textsc{IsDate}     & Value is a date
  & 11 & \textsc{\#Spaces}   & Space count
  & 18 & \textsc{\#Neighbors} & Non-empty 4-neighbor count \\

  & 5 & \textsc{IsFormula}  & Cell contains a formula
  & 12 & \textsc{\#Other}    & Other character count
  & 19 & \textsc{IsIsolated} & All 4-neighbors are empty \\

  & 6 & \textsc{IsInt}      & Integer value
  & 13 & \textsc{StartsAlpha} & First char.\ is alphabetic
  & 20 & \textsc{IsOds}      & File format is ODS \\

  & 7 & \textsc{IsFloat}    & Non-integer numeric value
  & 14 & \textsc{StartsNum}  & First char.\ is numeric
  & 21 & \textsc{IsExcel}    & File format is XLS/XLSX \\

  \midrule

  \multirow{5}{*}{\rotatebox{90}{\textbf{Group B}}}
  & 22    & \textsc{Bold}     & Font weight is bold
  & 27--29 & \textsc{FontColor} & Font color (R, G, B $\in[0,1]$)
  & 36     & \textsc{BorderTop}    & Has top border \\

  & 23    & \textsc{Italic}   & Font style is italic
  & 30--32 & \textsc{BgColor}   & Background col. (R, G, B $\in[0,1]$)
  & 37     & \textsc{BorderBottom} & Has bottom border \\

  & 24    & \textsc{Underline}& Text is underlined
  & 33     & \textsc{IsMerged}   & Cell belongs to a merged region
  & 38--44 & \textsc{HAlign}    & Horiz.\ alignment (7-class one-hot) \\

  & 25    & \textsc{FontSize} & Font size (pt)
  & 34     & \textsc{BorderLeft} & Has left border
  & 45--50 & \textsc{VAlign}    & Vert.\ alignment (6-class one-hot) \\

  & 26    & \textsc{FontHash} & Stable hash of font name
  & 35     & \textsc{BorderRight}& Has right border
  & & & \\

  \midrule

  \multirow{5}{*}{\rotatebox{90}{\textbf{Group C}}}
  & 51 & \textsc{RowNonEmptyRatio} & Fraction non-empty in row
  & 56 & \textsc{IsAloneInRow}     & Only non-empty cell in row
  & 61 & \textsc{IsAloneInCol}     & Only non-empty cell in column \\

  & 52 & \textsc{RowFracNum}    & Fraction numerical in row
  & 57 & \textsc{ColNonEmptyRatio} & Fraction non-empty in column
  & 62 & \textsc{ColTypeOutlier}   & Type differs from col.\ majority \\

  & 53 & \textsc{RowFracStr}    & Fraction string in row
  & 58 & \textsc{ColFracNum}       & Fraction numerical in column
  & 63 & \textsc{MergeColRatio}    & Fraction merged cells in column \\

  & 54 & \textsc{RowHomog}      & Type homogeneity in row
  & 59 & \textsc{ColFracStr}       & Fraction string in column
  & 64 & \textsc{MergeRowRatio}    & Fraction merged cells in row \\

  & 55 & \textsc{RowIsUniform}  & All non-empty row cells same type
  & 60 & \textsc{ColHomog}         & Type homogeneity in column
  & 65 & \textsc{MergeColSpan3+}   & Merge spans $\geq 3$ columns \\

  \bottomrule
  \end{tabular}
\end{table*}

\bparagraph{Classification Pipeline}\label{sec:ctc_pipeline}%
The classification pipeline addresses two complementary limitations of
cell-wise prediction. First, most Group A and B features are binary or
categorical (\textsc{Bold}, \textsc{IsNum}, \textsc{StartsAlpha}\ldots), and
class membership typically depends on their joint interactions rather
than individual signals;
as an example, \cls{HEADER} cells often combine %
formatting and row-level homogeneity cues that are not
linearly separable. This motivates a \emph{non-linear classifier}, which
captures feature conjunctions better than linear models,
as confirmed empirically in Sec.~\ref{sec:exp_ctc}.
Second, cell labels exhibit strong spatial dependencies in the grid:
a cell may be \cls{HEADER} partly because it precedes a \cls{DATA} region, and \cls{TITLE}
cells are often defined by their structural position relative to headers and
data blocks. To model these dependencies, we use a \emph{pairwise CRF}
to propagate information across adjacent cells and enforce neighborhood consistency.

\emph{Step 1: LightGBM classifier.}
We train a Gradient-Boosted Decision Tree (GBDT) classifier using LightGBM~\cite{ke2017lightgbm}
on the 65 unary features of all training cells. Compared to Random Forests (RF)~\cite{breiman2001random},
GBDT iteratively fits trees to correct residual errors of the current ensemble,
yielding stronger learners on heterogeneous feature sets combining binary flags, counts, ratios, and color values.
Class imbalance (\cls{DATA} and \cls{EMPTY} cells dominate most sheets) is handled
by reweighting classes inversely to their frequency in the loss.
The model outputs a per-class probability vector
$\boldsymbol{\pi}_i = \text{LightGBM}\left(\mathbf{x}_i^\text{u}\right) \in [0,1]^{|\mathcal{C}|}$ for each cell $i$, where $\mathbf{x}_i^\text{u}$
is the 65D feature vector.

\emph{Step 2: \textsc{EdgeFeatureGraphCRF}.}
Building on the grid $G$ defined above, we define a first \emph{CRF scoring} function as:
\begin{equation}
F^{\textsc{Linear}}(\mathbf{x}, \mathbf{y}) = \sum_{i = 1}^{|V|} {\mathbf{W}^\text{u}_{\left(y_i\right)}}^\top \mathbf{x}_i^\text{u} +
  \sum_{(i, j) \in E} {\mathbf{W}^\text{p}_{\left(y_i, y_j\right)}}^\top \mathbf{x}_{ij}^\text{p}
  \label{eq:crf_score}
\end{equation}
where $\mathbf{x}_i^\text{u} \in \mathbb{R}^{65}$ and
$\mathbf{x}_{ij}^\text{p} \in \mathbb{R}^{30}$ are the unary and pairwise
feature vectors, $\mathbf{W}^\text{u} \in \mathbb{R}^{65 \times |\mathcal{C}|}$ contains
one weight vector per class, and $\mathbf{W}^\text{p} \in \mathbb{R}^{30 \times |\mathcal{C}|^2}$ one %
per ordered label pair. We use the \textsc{EdgeFeatureGraphCRF} of~%
\cite{muller2014pystruct}, which learns a distinct pairwise weight vector $\mathbf{W}^\text{p}_{(k,l)}$ per label
pair $(k, l)$, allowing label-specific pairwise interactions rather than shared across transitions.
For instance, a \cls{HEADER}$\to$\cls{DATA} transition may be encouraged when neighboring cells differ in
alignment or font weight, and discouraged otherwise.
Eq.~\eqref{eq:crf_score} corresponds to a purely linear CRF using the 65 unary features.
To incorporate the non-linear LightGBM classifier, we %
replace the feature vector $\mathbf{x}^\text{u}_i$ with the class-probability
vector $\boldsymbol{\pi}_i$, yielding:
\vspace*{-4pt}
\begin{equation}
  F^{\textsc{LightGBM}}(\mathbf{x}, \mathbf{y}) = \sum_{i = 1}^{|V|} {\mathbf{W}^\pi_{\left(y_i\right)}}^\top \boldsymbol{\pi}_i +
  \sum_{(i, j) \in E} {\mathbf{W}^\text{p}_{\left(y_i, y_j\right)}}^\top \mathbf{x}_{ij}^\text{p}
  \label{eq:crf_score_probabilities}
\end{equation}
\vspace*{-1pt}%
\noindent where $\mathbf{W}^\pi \in \mathbb{R}^{|\mathcal{C}|\times|\mathcal{C}|}$. In Sec.~\ref{sec:exp_ctc},
we compare both cases experimentally: Eq.~\eqref{eq:crf_score} as \cls{CRF-Linear}, and
Eq.~\eqref{eq:crf_score_probabilities} as \cls{CRF-LightGBM}.

\emph{MAP (Maximum A Posteriori) inference.}
At inference, the labeling for a spreadsheet $\mathbf{x}$ is obtained by
maximizing the function~$F$.
Exact MAP inference over a general pairwise CRF is NP-hard~\cite{shimony1994finding}, so we use
the $\alpha$-expansion algorithm~\cite{boykov2001fast}, via the
\href{https://github.com/pystruct/pyqpbo}{\emph{PyQPBO}}
library. It is a move-making
algorithm that iteratively proposes a candidate label~$\alpha \in \mathcal{C}$ and, for each~$\alpha$, jointly optimizes over
all cells the binary decision of keeping the current label or switching
to $\alpha$. This reduces the multi-class problem to a sequence of binary
graph-cut subproblems until no move decreases the energy
(convergence). %
Each binary subproblem is solved using Quadratic
Pseudo-Boolean Optimization (QPBO)~\cite{rother2007optimizing}, which reformulates
binary energy minimization ($-F(\mathbf{x}, \tilde{\mathbf{y}})$) as a minimum graph cut: pairwise terms
penalizing label disagreement between neighbors (attractive terms) are solved
exactly in polynomial time~\cite{kolmogorov2004what}, while terms that encourage label disagreement
(repulsive terms) are handled approximately~\cite{rother2007optimizing}.
This yields exactly the MAP result when all pairwise interactions are attractive, and a
strong approximation otherwise, while remaining tractable for grids of
several thousand cells seen in~practice~\cite{kolmogorov2004what}.

\emph{CRF training.}
CRF parameters $\mathbf{W}$ ($\mathbf{W}^\text{u}$/$\mathbf{W}^\pi$ and $\mathbf{W}^\text{p}$) are learned by minimizing the
structured hinge loss via the structural SVM framework~\cite{tsochantaridis2005large}.
Given $n$ training sheets with ground-truth labelings $\{\mathbf{y}_t^*\}_{t=1}^n$, the idea is to find
$\mathbf{W}$ such that the score $F(\mathbf{x}_t, \mathbf{y}_t^*)$ exceeds the score of any competing
labeling $\bar{\mathbf{y}}_t$ by a margin proportional to how wrong $\bar{\mathbf{y}}_t$ is
(measured by the Hamming loss $\Delta(\mathbf{y}_t^*, \bar{\mathbf{y}}_t)$, i.e., the fraction of
mislabeled cells). A slack variable $\xi \geq 0$ absorbs margin violations when
the constraint cannot be satisfied exactly. In the $n$-slack variant, one slack
per training sheet is introduced, leading to a cutting-plane problem with
$O(n/\varepsilon^2)$ iterations to reach~$\varepsilon$-accuracy. The
one-slack formulation~\cite{joachims2009cutting} instead introduces a
\emph{single} slack variable summed over all training examples:
\[
  \begin{aligned}
    &\min_{\mathbf{W},\,\xi \geq 0}
    \;\frac{1}{2}\|\mathbf{W}\|^2 + \lambda\xi
    \quad \text{s.t.} \quad
    \forall\, \left(\bar{\mathbf{y}}_t\right)_{t=1}^{n} \in \prod_{t=1}^{n}\mathcal{C}^{|V_t|}\text{,}\\
    &\frac{1}{n}\sum_{t} \mathbf{W}^\top \bigl[\Psi(\mathbf{x}_t, \mathbf{y}_t^*) - \Psi(\mathbf{x}_t, \bar{\mathbf{y}}_t)\bigr]
    \geq \frac{1}{n}\sum_{t} \Delta(\mathbf{y}_t^*, \bar{\mathbf{y}}_t) - \xi
  \end{aligned}
\]
where $\Psi(\mathbf{x}_t, \mathbf{y})$ is the joint feature map (concatenation of
unary and pairwise sufficient statistics) and $\lambda > 0$ controls the
regularization strength. This constraint enforces the margin
\emph{on average} across all sheets rather than per sheet, reducing the
cutting-plane complexity to $O(1/\varepsilon)$ iterations regardless of $n$.
The trade-off is that per-sheet margin violations can be hidden by other
sheets, potentially allowing a small number of poorly labeled sheets to go
uncorrected. However, the $n$-slack variant does not scale to large datasets, and the
\emph{average} constraint remains sufficiently representative in our setting where
$n$ exceeds 500 sheets per fold (see Sec.~\ref{sec:dataset} and Sec.~\ref{sec:experiments} for
dataset information and experimental setup).

\emph{Handling class imbalance in the linear CRF baseline.}
For the linear CRF ablation (\cls{CRF-Linear}, Sec.~\ref{sec:exp_ctc}) using
raw unary features $\mathbf{x}^\text{u}_i$ instead of %
$\boldsymbol{\pi}_i$, class imbalance must be handled explicitly. We extend pystruct's
\textsc{EdgeFeatureGraphCRF} with a class-balanced structured hinge loss, weighting each
cell by the inverse square root of its class frequency within the training fold.
We additionally evaluate an optional \emph{hard-\cls{EMPTY} clamping} strategy: for cells with
\textsc{IsEmpty}=1, a large negative bias is applied to all non-\cls{EMPTY} unary scores,
effectively preventing the CRF from predicting other classes.
Analogously, we prevent prediction of \cls{EMPTY} when \textsc{IsEmpty}=0.
These mechanisms are unnecessary in the final \cls{CRF-LightGBM}, as class imbalance is already handled by
LightGBM, and
empty cells naturally receive near-zero values to non-\cls{EMPTY} classes.%

\subsection{Table Detection}\label{sec:approach_td}

For a spreadsheet region of $H$ rows and
$W$ columns, we have  $H\times W = |V|$  where $V$ is the cell
(vertex) set from Problem~\ref{problem:1}. Having obtained the cell type prediction
$\hat{\mathbf{y}}$ as discussed in Sec.~\ref{sec:approach_ctc},
let $\hat{G} \in \mathcal{C}^{H \times W}$ be the
predicted cell-type label grid. Formally:

\begin{problem}[Table Detection]
Given $\hat{G}$, the TD problem consists of predicting a set
  of non-overlapping, axis-aligned rectangles $\mathcal{T} = \{T_k\}$, where each $T_k$ is
  the minimal bounding box enclosing a maximal coherent group of \cls{HEADER} and
  \cls{DATA} cells forming a single table in $\hat{G}$.
\end{problem}

We solve this with a deterministic algorithm
that requires no learning.
It follows a five-stage pipeline mirroring
the heuristics a human annotator
would apply when delineating
tables on a grid of class labels:
(1) \emph{seeding} extracts maximal
connected regions of \cls{HEADER}/\cls{DATA} cells; (2) \emph{merging} fuses pairs of regions
separated only by within-table artifacts (sub-header rows, inter-table titles,
thin empty bands); (3) \emph{filtering} discards sparse or incoherent candidates; (4)
\emph{expansion} recovers locally disconnected top/left/right header bands; (5)
\emph{deduplication} resolves residual overlaps.
The method uses a small set of interpretable hyperparameters
with default values provided when introduced\versioning{ (summarized in~\cite{gauquier2026structuredextended}).}{. %
They are also summarized in Table~\ref{tab:td_hparams}.}

\emph{Notation and definitions.}
Rows are indexed by $r \in \{0, \dots, H-1\}$ and columns by
$c \in \{0, \dots, W-1\}$. We introduce:

\begin{definition}[Ranges and Densities]
A \emph{range} is a non-empty axis-aligned rectangle
$R = [r_0, r_1] \times [c_0, c_1]$ with area
$|R| = (r_1 - r_0 + 1)(c_1 - c_0 + 1)$. The
\emph{Intersection-over-Union} and the (asymmetric) \emph{containment ratio}
of two ranges are, respectively,
$
\mathrm{IoU}(R, R') = \frac{|R \cap R'|}{|R \cup R'|}
$,
$
\mathrm{cont}(R, R') = \frac{|R \cap R'|}{|R|}.
$
For a class subset $S \subseteq \mathcal{C}$, the \emph{$S$-density} of $R$ is
$
\rho_S(R) = \frac{1}{|R|} \sum_{(r, c) \in R} \mathds{1}\bigl[\hat{G}_{r,c} \in S\bigr].
$
For a row $r$ and a column span $[c_0, c_1]$, the \emph{absolute} and
\emph{relative} header densities are:
\(
\eta(r, c_0, c_1) =
\rho_{\left\{\text{HEADER}\right\}}\left(r\times\left[c_0, c_1\right]\right)
     $ and
$\eta^\star(r, c_0, c_1) =
\frac{\rho_{\left\{\text{HEADER}\right\}}\left(r\times\left[c_0, c_1\right]\right)}
     {\rho_{\mathcal{C}\setminus\left\{\text{EMPTY}\right\}}\left(r\times\left[c_0, c_1\right]\right)},
\)
with $\eta^\star = 0$ by convention when the denominator is 0.
The definitions of $\eta(c, r_0, r_1)$ and $\eta^\star(c, r_0, r_1)$ are
symmetric for columns.
\end{definition}
Above, the \emph{absolute} score $\eta$ measures how much of a column range is covered by
\cls{HEADER} cells. Empty cells reduce $\eta$, which is useful when deciding
whether a row separates two table candidates: a partial header should not span
both tables.
  The \emph{relative} score $\eta^\star$,
instead, measures the
proportion of \emph{non-empty} cells in a row that are headers, ignoring empty
padding. This is useful when deciding whether a sparse row acts as a header:
rows that span wider than the data block below them often contain blank cells
above groups of columns, so $\eta$ would be diluted by those empty cells even
for a genuine header row, whereas $\eta^\star$ remains high.

\begin{definition}[Seed Mask and Seeds]\label{def:seed}
  The \emph{seed mask} of $\hat{G}$ is the binary matrix $M \in \{0,1\}^{H \times W}$
  defined by $M_{r,c} = \mathds{1}\bigl[\hat{G}_{r,c} \in \{\cls{HEADER}, \cls{DATA}\}\bigr]$
  for all $(r, c)$.
  We define the graph $\left(V_M, E_M\right)$ as
  $V_M = \left\{(r,c) : M_{r,c} = 1\right\}$,
  $E_M = \left\{\{(r,c),(r',c')\} : |r-r'|+|c-c'| = 1\right\}$
  (horizontal/vertical adjacency).
  A \emph{seed} $S$ is a connected component of $\left(V_M, E_M\right)$. We derive
  its \emph{row} and \emph{column extents}
  $r_{\min}(S)$, $r_{\max}(S)$, $c_{\min}(S)$, $c_{\max}(S)$,
  and \emph{bounding box}
  $
    B(S) = [r_{\min}(S),\, r_{\max}(S)] \times [c_{\min}(S),\, c_{\max}(S)].
  $
  We write $n_D(S) = \sum_{(r,c)\in S} \mathds{1}[\hat{G}_{r,c}=\cls{DATA}]$
  and~$n_H(S)$ analogously for \cls{HEADER}.
  $\textsc{Seeds}(\hat{G})$ is the set of all seeds.
  \end{definition}

\definecolor{Dcolor}{HTML}{2A9D8F}
\definecolor{Hcolor}{HTML}{7570B3}
\definecolor{SeedBase}{HTML}{D95F02}

  \begin{figure}[t]
    \centering
    \scriptsize
    \setlength{\tabcolsep}{2pt}
    \renewcommand{\arraystretch}{1.3}
    $\hat{G}{=}$\,%
    \begin{tabular}[c]{r|ccccc}
      & 0 & 1 & 2 & 3 & 4 \\
    \hline
    0 & \cellcolor{gray!25}$T$ & \cellcolor{gray!25}$T$ & \cellcolor{gray!25}$T$ & \cellcolor{gray!25}$T$ & \cellcolor{gray!25}$T$ \\
    1 & \cellcolor{Hcolor}$H$ & \cellcolor{Hcolor}$H$ & \cellcolor{Hcolor}$H$ & \cellcolor{Hcolor}$H$ & \cellcolor{Hcolor}$H$ \\
    2 & \cellcolor{Hcolor}$H$ & \cellcolor{Dcolor}$D$ & \cellcolor{Dcolor}$D$ & \cellcolor{Dcolor}$D$ & \cellcolor{Dcolor}$D$ \\
    3 & $E$ & $E$ & $E$ & $E$ & $E$ \\
    4 & $E$ & \cellcolor{Hcolor}$H$ & \cellcolor{Hcolor}$H$ & \cellcolor{Hcolor}$H$ & \cellcolor{Hcolor}$H$ \\
    5 & \cellcolor{Hcolor}$H$ & \cellcolor{Dcolor}$D$ & \cellcolor{Dcolor}$D$ & \cellcolor{Dcolor}$D$ & \cellcolor{Dcolor}$D$ \\
    6 & $E$ & $E$ & $E$ & $E$ & $E$ \\
    7 & \cellcolor{gray!25}$T$ & \cellcolor{gray!25}$T$ & \cellcolor{gray!25}$T$ & \cellcolor{gray!25}$T$ & \cellcolor{gray!25}$T$ \\
    8 & \cellcolor{Hcolor}$H$ & \cellcolor{Hcolor}$H$ & \cellcolor{Hcolor}$H$ & \cellcolor{Hcolor}$H$ & \cellcolor{Hcolor}$H$  \\
    9 & \cellcolor{Dcolor}$D$ & \cellcolor{Dcolor}$D$ & \cellcolor{Dcolor}$D$ & \cellcolor{Dcolor}$D$  & \cellcolor{Dcolor}$D$ \\
    \hline
    \end{tabular}%
    \,${\to}\,M{=}$\,%
    \begin{tabular}[c]{r|ccccc|l}%
       & 0 & 1 & 2 & 3 & 4 \\
    \hline
    0 & \cellcolor{gray!25}0 & \cellcolor{gray!25}0 & \cellcolor{gray!25}0 & \cellcolor{gray!25}0 & \cellcolor{gray!25}0 & {\color{gray}TITLE --- not in mask} \\
    1 & \cellcolor{SeedBase}1 & \cellcolor{SeedBase}1 & \cellcolor{SeedBase}1 & \cellcolor{SeedBase}1 & \cellcolor{SeedBase}1 &
      \multirow{2}{*}{$\big\}$ Seed $a$; $B(a) = [1, 2] \times [0, 4]$} \\
    2 & \cellcolor{SeedBase}1 & \cellcolor{SeedBase}1 & \cellcolor{SeedBase}1 & \cellcolor{SeedBase}1 & \cellcolor{SeedBase}1 \\
    3 & 0 & 0 & 0 & 0 & 0 & {\color{gray}1 separator (sep.) --- isolated EMPTY} \\
    4 & 0 & \cellcolor{SeedBase}1 & \cellcolor{SeedBase}1 & \cellcolor{SeedBase}1 & \cellcolor{SeedBase}1 &
    \multirow{2}{*}{\begin{minipage}[t]{4cm}%
        Seed $b$; $B(b) = [4, 5]\times[0, 4]$ \\ $\tau=\eta_4=0.8\geq\tau_H$ $\land$ 1~sep.$\Rightarrow$ \textbf{merge ($a$, $b$)}\end{minipage}} \\
    5 & \cellcolor{SeedBase}1 & \cellcolor{SeedBase}1 & \cellcolor{SeedBase}1 & \cellcolor{SeedBase}1 & \cellcolor{SeedBase}1 & \\
    \hline
    6 & 0 & 0 & 0 & 0 & 0 &
      \multirow{2}{*}{
        $\big\}$ \color{gray} 2 separators (2 sep.) --- EMPTY then TITLE} \\
    7 & \cellcolor{gray!25}0 & \cellcolor{gray!25}0 & \cellcolor{gray!25}0 & \cellcolor{gray!25}0 & \cellcolor{gray!25}0 & \\
    8 & \cellcolor{SeedBase}1 & \cellcolor{SeedBase}1 & \cellcolor{SeedBase}1 & \cellcolor{SeedBase}1 & \cellcolor{SeedBase}1 &
      \multirow{2}{*}{\begin{minipage}[t]{4cm}%
        Seed $c$; $B(c) = [8, 9]\times[0, 4]$ \\ $\tau=\eta_8=1\geq\tau_H \land$ 2~sep. $>1 \Rightarrow$ \textbf{no merge}
      \end{minipage}} \\
    9 & \cellcolor{SeedBase}1 & \cellcolor{SeedBase}1 & \cellcolor{SeedBase}1 & \cellcolor{SeedBase}1 & \cellcolor{SeedBase}1 & \\
    \hline
    \end{tabular}
    \vspace*{-1em}
    \caption{Seeding and merging on a two-table grid $\hat{G}$
    ({\color{gray!25}$\blacksquare$}~\cls{TITLE},
    {\color{Hcolor}$\blacksquare$}~\cls{HEADER},
    {\color{Dcolor}$\blacksquare$}~\cls{DATA}, white~\cls{EMPTY}).
    $M_{r,c}=\mathds{1}[\hat{G}_{r,c}\in\{H,D\}]$ yields seeds $a$, $b$, $c$.
    $\eta_r$ abbreviates $\eta(r,0,4)$.}
    \label{fig:seeding_example}
    \end{figure}

The full detection procedure $\textsc{TableRangeExtractor}(\hat{G})$ is given in Algorithm~\ref{alg:td_main}:
starting from a set of seeds, it repeatedly merges compatible candidates, filters incoherent ones, expands
the remaining candidates into table ranges, and removes duplicates. %

\begin{algorithm}[t]
\caption{$\textsc{TableRangeExtractor}(\hat{G})$}
\label{alg:td_main}
\begin{algorithmic}[1]
\Require Cell-type grid $\hat{G} \in \mathcal{C}^{H \times W}$
\Ensure Set of detected table ranges $\mathcal{T}$
\State \label{lin:seeds}$\mathcal{S} \gets \textsc{Seeds}(\hat{G})$
\State $\mathcal{S} \gets \left\{S \in \mathcal{S} : n_D(S) \geq n_D^{\min}
  \lor \left(n_H({S}) \geq n_H^{\min} \land n_D(S) = 0\right) \right\}$
\While{$\exists \left(S_i, S_j\right) \in \mathcal{S}^2$, $i \neq j$,
s.t. \(\textsc{ShouldMerge}(S_i, S_j, \raisebox{0pt}[0pt][0pt]{$\hat G$})\)}
  \label{lin:mergebegin}
  \State Pick any such pair $\left(S_i, S_j\right)$
  \State \label{lin:mergeend}$\mathcal{S} \gets \left(\mathcal{S} \setminus \{{S}_i, {S}_j\}\right)
  \cup \left\{{S}_i \cup {S}_j\right\}$
\EndWhile
\State $\mathcal{S} \gets \bigl\{{S} \in \mathcal{S} \;:\; \textsc{Coherent}({S}, \hat{G})\bigr\}$
\State $\mathcal{T} \gets \bigl\{\textsc{Expand}({S}, \hat{G}) \;:\; {S} \in \mathcal{S}\bigr\}$
\State \Return $\textsc{Deduplicate}(\mathcal{T})$
\end{algorithmic}
\end{algorithm}

\emph{Stage 1: Seeding.}
We define a table as the bounding box covering its \cls{HEADER} and \cls{DATA} cells
(excluding \cls{TITLE} and \cls{OTHER} cells). We compute the
connected components of the graph $\left(V_M, E_M\right)$; %
each yields a candidate seed (line~\ref{lin:seeds} of Algorithm~\ref{alg:td_main}).
We retain only seeds with $n_D \geq n_D^{\min} = 4$
data cells, or header-only seeds ($n_D = 0$) with $n_H \geq n_H^{\min} = 8$:
the latter preserves genuine header bands disconnected from their data block
by over-segmentation; such seeds can later be recovered during the merging stage.
Figure~\ref{fig:seeding_example} illustrates the seeding on a two-table grid,
yielding three seeds.

\emph{Stage 2: Merging.}
Connected components in the raw seed mask are \textbf{systematically over-segmented} for two reasons.
\textbf{1.} Tables may contain \emph{sub-header rows} splitting the data region
(e.g., section dividers between row groups).
These rows are correctly classified as
\cls{HEADER}, but their column-wise structure may differ from that of the top header,
leaving them disconnected from rows above and below, splitting a single
table into multiple seeds.
\textbf{2.} Adjacent tables on the same sheet are often separated by
a thin band of \cls{EMPTY} rows together with a \cls{TITLE} row introducing the next table;
on the seed mask this appears as empty space, which does not reliably
distinguish \emph{inter-table separation} from \emph{intra-table structure}.
Figure~\ref{fig:seeding_example} illustrates both: seeds $a$ and $b$ are
separated by an empty row, but row~4 acts as a sub-header sufficiently overlapping both
column structures, thus triggering a merge. In contrast, seeds $b$ and $c$ are separated by an
empty row \emph{and} a title row; this blocks merging.
We address both cases (\textbf{1.} and \textbf{2.}) via the predicate \textsc{ShouldMerge}
(Algorithm~\ref{alg:td_merge}), applied iteratively until a fixed point
(lines~\ref{lin:mergebegin}--\ref{lin:mergeend} of
Algorithm~\ref{alg:td_main}); each round strictly decreases
$|\mathcal{S}|$, ensuring convergence. The \emph{vertical} case
(lines~\ref{lin:vbegin}--\ref{lin:vend} of Algorithm~\ref{alg:td_merge}) handles seeds with overlapping
column spans separated by a vertical gap. Along the gap, an intermediate row
with header density $\eta \geq \tau_H = 0.4$ over the shared column range is
read as an interior sub-header and triggers immediate merge; otherwise, the
decision combines the lower seed's first-row density with the number of
separator rows (containing only \cls{EMPTY}/\cls{TITLE} cells over the shared
span), reflecting that intra-table interruptions are typically shorter than
inter-table whitespace ($g_v = 4$). The \emph{horizontal} case
(lines~\ref{lin:hbegin}--\ref{lin:hend}) is symmetric and simpler: seeds are merged when fewer than
$g_h = 2$ fully empty columns separate them. This suffices because horizontal
gaps lack the sub-header and title artifacts that complicate the vertical~case.

\begin{algorithm}[t]
\caption{$\textsc{ShouldMerge}$: merge predicate.}
\label{alg:td_merge}
\begin{algorithmic}[1]
\Require Seeds ${S}_i, {S}_j$; grid $\hat{G}$
\Ensure Boolean indicating whether to merge ${S}_i$ and ${S}_j$
\State \label{lin:vbegin}$c_l \gets \max\left(c_{\min}({S}_i), c_{\min}({S}_j)\right)$;
       $c_r \gets \min\left(c_{\max}({S}_i), c_{\max}({S}_j)\right)$
\If{$c_l \leq c_r$}
  \State $r_t$$\gets$$\min\bigl(r_{\max}({S}_i), r_{\max}({S}_j)\bigr)$; $r_b$$\gets$$\max\bigl(r_{\min}({S}_i), r_{\min}({S}_j)\bigr)$
  \If{$r_b > r_t + 1$}
    \State $\mathrm{sep} \gets 0$
    \For{$r$ from $r_t + 1$ to $r_b - 1$}
      \State \textbf{if} $\eta\left(r, c_l, c_r\right) \geq \tau_H$ \textbf{then} \textbf{return true}
      \If{$\hat{G}_{r,c} \in \{\cls{EMPTY}, \cls{TITLE}\}$ for every $c \in [c_l, c_r]$}
        \State $\mathrm{sep} \gets \mathrm{sep} + 1$
      \EndIf
    \EndFor
    \State $\tau \gets \eta\left(r_b, c_l, c_r\right)$
    \State \label{lin:vend}\Return $\bigl(\tau \geq \tau_H \land \mathrm{sep} \leq 1\bigr)\;\lor\;\bigl(\tau < \tau_H \land \mathrm{sep} < g_v\bigr)$
  \EndIf
\EndIf
\State \label{lin:hbegin}$r_l \gets \max\bigl(r_{\min}({S}_i), r_{\min}({S}_j)\bigr)$;
       $r_r \gets \min\bigl(r_{\max}({S}_i), r_{\max}({S}_j)\bigr)$
\If{$r_l \leq r_r$}
  \State $c_t$$\gets$$\min\bigl(c_{\max}({S}_i), c_{\max}({S}_j)\bigr)$; $c_b$$\gets$$\max\bigl(c_{\min}({S}_i), c_{\min}({S}_j)\bigr)$
  \If{$c_b > c_t + 1$}
    \State $\mathrm{ev} \gets \lvert\{c \in [c_t+1,c_b-1]:\forall r \in [r_l,r_r], \hat{G}_{r,c}=\cls{EMPTY}\}\rvert$
    \State \label{lin:hend}\Return $\bigl(\mathrm{ev} < g_h\bigr)$
  \EndIf
\EndIf
\State \Return \textbf{false}
\end{algorithmic}
\end{algorithm}

\emph{Stage 3: Coherence filtering.}
Merging may still produce two types of \textbf{invalid candidates}: ($i$)~
large, sparse seeds formed by repeated merges connecting distant regions via
narrow bridges of \cls{HEADER} or \cls{DATA} cells---large bounding boxes with
little table content;
($ii$)~data blocks whose header was misclassified,
surviving seeding without corresponding to a valid table.
The predicate
$\textsc{Coherent}({S}, \hat{G})$ handles both cases.
It rejects a seed if both densities in $B(S)$ fall
below thresholds, i.e.,
$\rho_{\{\cls{DATA}\}}(B(S)) < \rho_D^{\min} $$=$$ 0.05$ \emph{and}
$\rho_{\{\cls{HEADER}\}}(B(S))$ $< \rho_H^{\min} = 0.5$.
Data-only seeds are rejected when they contain fewer
than $n_D^{\dagger} = 20$ \cls{DATA} cells and have no \cls{HEADER} cell within
Chebyshev radius $r_N$$=$$3$ of $B({S})$. Header-only seeds %
are
retained iff $n_H \geq n_H^{\min}$.%

\emph{Stage 4: Expansion.}
By construction, a surviving seed ${S}$
contains the data block of a table and header cells 4-connected to it in the seed mask.
However, \textbf{some header cells
may remain disconnected} because of empty padding rows, merged-cell rendering artifacts, or isolated
cells misclassified as \cls{TITLE} or \cls{OTHER}. %
We recover them via a directional expansion procedure \printIfExtVersion{(Algorithm~\ref{alg:td_expand})}{(Algorithm in~\cite{gauquier2026structuredextended})} that grows
the seed bounding box upward, leftward, and rightward, but not downward, since headers
in statistical spreadsheets conventionally appear above or beside the data, never below.
Expansion continues while the next row (resp.\ column) has relative header
density $\eta^\star \geq \tau_H$ (resp.\ $\tau_H^c$), two thresholds set to $0.4$.
We use $\eta^\star$ rather than $\eta$ because header rows often contain empty padding above
column groups, where $\eta$ would be diluted by those blanks even on genuine
headers. Column scans \printIfExtVersion{(lines~\ref{lin:colleft} and~\ref{lin:colright})}{} are evaluated over the seed's
\emph{original} row range, so side-header columns remain anchored to the
original data region rather than to newly recovered top headers.

\printIfExtVersion{%
\begin{algorithm}[t]
\caption{$\textsc{Expand}$: expansion of a seed into a table range.}
\label{alg:td_expand}
\begin{algorithmic}[1]
\Require Seed ${S}$, grid $\hat{G}$
\Ensure Range $R$ representing the detected table
\State $\left(\bar r_0, \bar r_1, \bar c_0, \bar c_1\right) \gets
        \left(r_{\min}({S}), r_{\max}({S}),
        c_{\min}({S}), c_{\max}({S})\right)$
\State $r_0 \gets \bar r_0$;\quad $r \gets \bar r_0 - 1$
\State \textbf{while} $r \geq 0$ \textbf{and} $\eta^\star(r, \bar c_0, \bar c_1) \geq \tau_H$ \textbf{do} $\bigl(r_0 \gets r$; $r \gets r - 1\bigr)$
\State $c_0 \gets \bar c_0$;\quad $c \gets \bar c_0 - 1$
\State \label{lin:colleft}\textbf{while} $c \geq 0$ \textbf{and} $\eta^\star(c, \bar r_0, \bar r_1) \geq \tau_H^c$ \textbf{do} $\bigl(c_0 \gets c$; $c \gets c - 1\bigr)$
\State $c_1 \gets \bar c_1$;\quad $c \gets \bar c_1 + 1$
\State \label{lin:colright}\textbf{while} $c < W$ \textbf{and} $\eta^\star(c, \bar r_0, \bar r_1) \geq \tau_H^c$ \textbf{do} $\bigl(c_1 \gets c$; $c \gets c + 1\bigr)$
\State \Return $[r_0, \bar r_1] \times [c_0, c_1]$
\end{algorithmic}
\end{algorithm}}{}

\emph{Stage 5: Deduplication.}
In rare cases, \textbf{two distinct seeds yield overlapping table ranges},
e.g., when a single header band spans two nearby column-aligned data blocks.
We resolve these with a Non-Maximum-Suppression (NMS) step: ranges are
processed in decreasing order of area, and a range $R$ is accepted unless an
already-accepted range $R'$ satisfies
$\mathrm{IoU}(R, R') \geq \tau_{\mathrm{IoU}} = 0.5$ (strong overlap) or
$\mathrm{cont}(R, R') \geq \tau_{\mathrm{cont}} = 0.8$ (near-containment). The
remaining ranges form the final detection set $\mathcal{T}$.

\printIfExtVersion{%
\begin{table}[t]
\caption{Hyperparameters of the table range detector.}
\vspace*{-1em}
\label{tab:td_hparams}
\centering
\small
\begin{tabular}{lcll}
\toprule
 HP & Default & Stage & Description \\
\midrule
$n_D^{\min}$            & $4$    & 1              & Min.\ \cls{DATA} cells in a seed \\
$n_H^{\min}$            & $8$    & 1,3         & Min.\ \cls{HEADER} cells in a header-only seed \\
$\tau_H$                & $0.4$  & 2,4    & Header-density threshold along rows \\
$\tau_H^c$              & $0.4$  & 4            & Header-density threshold along columns \\
$g_v$                   & $4$    & 2              & Min.\ separator rows to keep two seeds apart \\
$g_h$                   & $2$    & 2              & Min.\ \cls{EMPTY} columns to keep two seeds apart \\
$\rho_D^{\min}$         & $0.05$ & 3            & Min.\ \cls{DATA} density of a seed bounding box \\
$\rho_H^{\min}$         & $0.5$  & 3            & Min.\ \cls{HEADER} density of a seed bounding box \\
$n_D^{\dagger}$         & $20$   & 3            & Acceptance threshold for header-less seeds \\
$r_N$                   & $3$    & 3            & Header-neighborhood radius (Chebyshev) \\
$\tau_{\mathrm{IoU}}$   & $0.5$  & 5        & IoU threshold for NMS \\
$\tau_{\mathrm{cont}}$  & $0.8$  & 5        & Containment threshold for NMS \\
\bottomrule
\end{tabular}
\end{table}}{}

\section{Dataset}
\label{sec:dataset}

Given the limitations of existing datasets for CTC and TD (Sec.~\ref{sec:related_work}),
we construct and release \datasetname{}~\cite{github_repository}, a new dataset of
spreadsheets annotated for both tasks, originating from statistical datasets published by
public organizations.

\bparagraph{Construction}
The dataset is highly heterogeneous, with respect to:
\begin{inparaenum}[(i)] \item the data sources,
\item the language used in the sheets, \item the covered topics, \item the sheet sizes,
and \item the file formats. \end{inparaenum}
This enables evaluating the robustness of cell-type classifiers and table detectors in highly
variable settings, which is critical for downstream applications relying on clean and structured
tables, such as %
fact-checking~\cite{gauquier2026efficientandscalable}.

\versioning{We consider 14 selected sources for the dataset (detailed composition in the extended version~\cite{gauquier2026structuredextended}),}{Table~\ref{tab:sources} shows the distribution of sheets across the 14 selected sources,}
chosen for
their geographical diversity,
which also increases language diversity. Topic diversity results from the broad scope of the
selected sources.
The collection proceeds as follows. We first retrieve a large pool of spreadsheets from each
source using a state-of-the-art focused Web crawler~\cite{gauquier2026efficientcrawling}. Each spreadsheet is then processed to identify candidate sheets
likely to contain statistical tables using a simple but effective heuristic: a sheet is retained if
at least 50\% of its non-empty cells are either purely numerical %
or predominantly numerical (i.e., a majority of characters are digits, allowing symbols or
precision annotations).
To limit source imbalance, we retain at most 100 candidate sheets per source and only one sheet
per spreadsheet file. %
The resulting candidates are randomly shuffled and
manually annotated, keeping only %
sheets containing at least one statistical table. %
Annotation continues until reaching 750 sheets, corresponding to $\sim$150
test sheets under the 5-fold cross-validation setup of Sec.~\ref{sec:experiments}.
A second sanitization pass then %
corrects annotation errors and removes ambiguous cases, yielding a final dataset of \textbf{737 annotated sheets},
\printIfExtVersion{\begin{table}
	\centering
	\caption{Distribution of the sheets per source of origin}
    \vspace*{-1em}
	\begin{tabular}{llr}
		\toprule
			\bfseries Source & \bfseries Country & \bfseries \#Sheets \\
		\midrule
            Australian Bureau of Statistics       & Australia       & 64 \\
            French Ministry of Justice                   & France          & 64 \\
            OECD                                         & International   & 63 \\
            Nat.\ Center for Education Statistics & USA             & 61 \\
            Bureau of Economic Analysis            & USA             & 58 \\
            Ministry of Internal Affairs \& Comm.\  & Japan           & 58 \\
            INSEE                                        & France          & 57 \\
            General Authority for Statistics    & Saudi Arabia    & 57 \\
            United States Census Bureau                  & USA             & 54 \\
            World Bank                                   & International   & 54 \\
            Ministry of Interior                         & France          & 50 \\
            Intl.\ Labour Organization             & International   & 48 \\
            World Health Organization              & International   & 48 \\
            CNIS                                         & France          &  1 \\
        \bottomrule
	\end{tabular}
	\label{tab:sources}
\end{table}}{}
written in six languages: English, French, Arabic, Japanese,
Spanish, and Icelandic.
The file formats are: XLSX (316 sheets, 42.9\%), XLS (296, 40.2\%),
ODS (60, 8.1\%), CSV (52, 7.1\%), and TSV (13, 1.8\%). Rich spreadsheet formats
(XLS, XLSX, ODS) provide formatting information; plain-text formats (CSV, TSV) do not.

\bparagraph{Annotation Protocol and Statistics}
Each retained sheet is annotated for both the CTC and TD tasks.
For
\textbf{CTC}, %
we annotate each
cell within the minimal bounding rectangle containing all the
non-empty cells of the sheet. We use the
five labels: \cls{EMPTY}~(0), \cls{HEADER}~(1), \cls{DATA}~(2), \cls{TITLE}~(3), and \cls{OTHER}~(4).
\cls{HEADER} includes column/row headers and sub-headers, which
may be adjacent to higher-level headers or appear
between \cls{DATA} cells.
Each \cls{DATA} cell is associated with at least one (and typically several) \cls{HEADER} cell.
\cls{TITLE} cells describe table content and are usually, though not necessarily,
located above it. \cls{OTHER} covers remaining non-empty cells %
(e.g., footnotes, annotations, data sources, measurement details), %
acting as a reject class. %
Overall, \textbf{7\,570\,354 cells} are annotated for CTC, with an average of 10\,272 per sheet.
Sheet sizes vary considerably, from 27 to 1\,542\,303 cells, illustrating strong structural variability in real-world spreadsheets.
The largest sheet is particularly useful for scalability evaluation %
(Sec.~\ref{sec:exp_td}). %
Class distribution is heavily imbalanced: \cls{DATA} cells account
for 62.2\% of the annotations, followed by \cls{HEADER} (20.4\%) and \cls{EMPTY} (17.1\%), while \cls{TITLE}
($<$0.1\%) and \cls{OTHER} ($<$0.2\%) are rare.

For the \textbf{TD} task, we annotate \emph{table ranges}.
Since each \cls{DATA}
cell is associated with at least one \cls{HEADER}, a table is defined as the rectangular bounding box
containing all \cls{HEADER} and \cls{DATA} cells belonging to it. %
Consequently, table ranges
contain only \cls{HEADER} and \cls{DATA} cells, excluding \cls{TITLE} and \cls{OTHER}. Moreover, each
\cls{HEADER} or \cls{DATA} cell can belong to only one table, meaning that overlapping tables are not
allowed. In practice, if two regions share \cls{HEADER} or \cls{DATA} cells, they are considered part of
the same table, potentially with hierarchical headers. These constraints %
are used as automatic sanitization checks to detect annotation inconsistencies before a manual
sanitization.
After annotation, sequences of more than two consecutive fully empty rows or columns are compressed to
exactly two, to save space and without losing information. %

Our 737 sheets lead to \textbf{818 annotated tables}. Most sheets (690) contain a single table; 29 contain two, and 18 contain
three or more.

\section{Experiments}
\label{sec:experiments}

Secs.~\ref{sec:exp_ctc} and \ref{sec:exp_td} respectively present the
CTC and TD experiments. %
All systems are evaluated  using 5-fold cross-validation: %
sheets are shuffled with a fixed random seed and split into five folds of
 $\approx$147 sheets each, yielding an 80/20 train/test split per fold.
Depending on the system, the training fold is used for
training, fine-tuning, hyperparameter selection, or not used.
The test fold is
shared across tasks for consistency. %
When a system
fails on some sheets, we report results on the subset successfully processed by all systems.

\subsection{Cell-Type Classification}
\label{sec:exp_ctc}

We compare several CTC systems, classified %
in Table~\ref{tab:ctc_systems} by their inclusion of spatial awareness
(\textbf{Spat.}), sequential modeling (\textbf{Seq.}), and non-linearity (\textbf{NL}). \printIfExtVersion{}{Hyperparameter choices are discussed in the extended version~\cite{gauquier2026structuredextended}.}

\begin{table}
    \centering
    \small
    \caption{Main characteristics of the CTC systems.}
    \vspace{-1em}
    \setlength{\tabcolsep}{5pt}
        \begin{tabular}{lcccccc}
        \toprule
        \textbf{Model} & \textbf{NL} & \textbf{Seq.} & \textbf{Fmt.} & \textbf{Spat.} & \textbf{Sem.} & \textbf{GPU} \\
        \midrule
        \cls{CRF-Linear}   & \xmark & \cmark & \cmark & \cmark & \xmark & \xmark \\
        \cls{RF-Koci}      & \cmark & \xmark & \cmark & \xmark & \xmark & \xmark \\
        \cls{RF}           & \cmark & \xmark & \cmark & \cmark & \xmark & \xmark \\
        \cls{LightGBM}     & \cmark & \xmark & \cmark & \cmark & \xmark & \xmark \\
        \cls{CRF-RF}       & \cmark & \cmark & \cmark & \cmark & \xmark & \xmark \\
        \cls{CRF-LightGBM} & \cmark & \cmark & \cmark & \cmark & \xmark & \xmark \\
        \cls{TUTA}         & \cmark & \cmark & \cmark & \cmark & \cmark & \cmark \\
        \bottomrule
        \end{tabular}
    \label{tab:ctc_systems}
    \end{table}

\noindent\textbf{\cls{CRF-Linear}} is the CRF model described in \emph{Step 2} of Sec.~\ref{sec:ctc_pipeline}.
It uses $\mathbf{x}^\text{u}$ and $\mathbf{x}^\text{p}$ for the unary and pairwise components, and thus
follows the score function of Eq.~\eqref{eq:crf_score}. The model includes the class-balanced loss and
\cls{EMPTY}-case clamping described in Sec.~\ref{sec:ctc_pipeline}. %
This method has sequential and spatial modeling, but is linear.
\printIfExtVersion{Hyperparameters are selected through a global 5-fold validation procedure to obtain a single configuration for all folds.
The configuration uses  $\lambda=0.1$, clamping, a square-root inverse-frequency class-balanced hinge loss, and a batch size of 128. Early stopping is based on the M-FM-F1 score (presented below), evaluated every 100 iterations.}{}

\noindent\textbf{\cls{RF-Koci}} is a baseline based on \emph{Step 1} of Sec.~\ref{sec:ctc_pipeline}, where LightGBM is replaced by a
RF classifier. It does not use the full unary feature set, but only the non-spatial
features, excluding all Group C features as well as feature \#19 from Group A. We retain feature \#18, the
only neighborhood-aware feature from~\cite{koci2016machine}. This baseline is therefore
a non-linear approach comparable to~\cite{koci2016machine},
without sequential modeling, and with minimal spatial awareness.
\printIfExtVersion{Hyperparameters are selected using the same shared 5-fold  validation procedure as for \cls{CRF-Linear}.
The retained hyperparameters include bootstrap sampling, a maximum depth
of 20, \texttt{log2} feature selection, a minimum of 7 samples per leaf, a minimum split size of 17, and 481 trees.}{}\printIfExtVersion{\par}{}
\noindent\textbf{\cls{RF}} is similar to \cls{RF-Koci} but uses the complete unary feature set ($\mathbf{x}^\text{u}$), thus it is non-linear.
\printIfExtVersion{Hyperparameters are selected using the same shared 5-fold validation procedure.
The retained configuration uses no bootstrap sampling, a maximum depth of
20, \texttt{sqrt} feature selection, a minimum of 6 samples per
leaf, a minimum split size of 18, and 574 trees.}{}

\noindent\textbf{\cls{LightGBM}} is a baseline following \emph{Step 1} of Sec.~\ref{sec:ctc_pipeline}, based on optimized GBDT. Like \cls{RF},
it is non-linear, models spatial awareness, but without sequential modeling.
\printIfExtVersion{Hyperparameters are selected using the same shared 5-fold validation procedure. The retained configuration uses a column subsampling rate of 0.6,
a learning rate of 0.06, a maximum depth of 8, 51 minimum child samples, 723 estimators, 111 leaves, $\ell_1$ and $\ell_2$ regularization of 0.01 and 0.5,
respectively, and a subsampling rate of 0.}{}

\noindent\textbf{\cls{CRF-RF}} applies a CRF on top of the \cls{RF} baseline (Sec.~\ref{sec:ctc_pipeline}), with LightGBM replaced
by RF (non-linear, sequential, with spatial awareness). We train the \cls{RF} model
and the CRF sequentially (the CRF using \cls{RF} probabilities $\boldsymbol{\pi}$ as unary inputs, following Eq.~\eqref{eq:crf_score_probabilities}) on each training fold, keeping the \cls{RF} hyperparameters
fixed and selecting CRF hyperparameters via the same 5-fold validation procedure.
\printIfExtVersion{The retained CRF configuration uses $\lambda=10$, no class-balanced loss,
no clamping, and a batch size of 128.}{}

\noindent\textbf{\cls{CRF-LightGBM}} combines a CRF model on top of \cls{LightGBM}, similarly to \cls{CRF-RF}. The CRF scoring function thus follows Eq.~\eqref{eq:crf_score_probabilities}.
\printIfExtVersion{It is therefore a non-linear (\textbf{NL}) and  sequential (\textbf{Seq.}) model with spatial awareness (\textbf{Spat.}). The same training procedure is used, keeping the
\cls{LightGBM} hyperparameters fixed and reusing the same CRF configuration as \cls{CRF-RF}.}{}

\noindent\textbf{\cls{TUTA}}~\cite{wang2021tuta}
(Sec.~\ref{sec:related_work})
 is pre-trained on large collections of unlabeled Web and
spreadsheet
\printIfExtVersion{tables using three progressive objectives at the token, cell, and table levels. The authors fine-tune \cls{TUTA}}
{tables, then fine-tuned }
on CTC
(which we focus on) and table type classification (TTC), which assumes known table boundaries, i.e., the output of
our TD task.
\cls{TUTA} is the only DL baseline we consider; it outperforms other DL-based CTC methods~\cite{wang2021tuta}. %
We use the official implementation~\cite{tuta_github}, adapted to our label set and
\printIfExtVersion{extended with a preprocessing pipeline to convert raw spreadsheets into the expected model inputs.}
{pipeline. }
We fine-tune the last two layers of the 12-layer backbone and the
classification head, reinitializing
from the pre-trained checkpoint at each fold. %
\printIfExtVersion{The original setup fine-tunes for 4 epochs on
  $\sim$3.5k tables ($\sim$1M labeled cells), while our dataset contains
  818 labeled tables with over 7M labeled cells;
we therefore use 20 epochs to match a comparable number of seen tables. We train with a learning rate of $3 \times 10^{-5}$ and batch size of 16, compared to $8 \times 10^{-6}$ and
batch size of 4 in the original work, following the linear scaling rule.}{}

\bparagraph{CTC Metrics}
The five cell types (recall $\mathcal{C}$ from Sec.~\ref{sec:approach_ctc}) %
are heavily imbalanced across sheets:
\cls{DATA} cells usually dominate, while \cls{TITLE} and
\cls{OTHER} cells are sparse and unevenly distributed.
Standard cell-level accuracy is therefore uninformative, and
cell-level macro F1 is distorted by sheet size as larger sheets
weigh more in the aggregate score.

Instead, we measure \textbf{per-class file-macro F1} (FM-F1),
which treats each sheet as a single observation.
Let $\mathcal{F}$ be the test sheets and $\mathcal{C}$ the
classes.
For $f \in \mathcal{F}$ and $c \in \mathcal{C}$,
let $F_1(f, c)$ denote the F1 score of class $c$ on sheet $f$.
Denoting $\mathcal{F}_c \subseteq \mathcal{F}$ the set of sheets
in which class $c$ is present (in ground truth or predictions),
the per-class file-macro F1 is
\(
    \text{FM-F1}_c = \frac{1}{|\mathcal{F}_c|} \sum_{f \in \mathcal{F}_c} F_1(f, c)
\).
Absent classes from a sheet are excluded from FM-F1$_c$,
avoiding both trivially perfect scores and unfair penalization of
rare classes. We summarize system performance with the \textbf{mean file-macro F1}
(M-FM-F1), the macro-average of FM-F1$_c$ across all classes
\(
    \text{M-FM-F1} = \frac{1}{|\mathcal{C}|} \sum_{c \in \mathcal{C}} \text{FM-F1}_c
\).

\bparagraph{CTC Computational Efficiency}
All systems except \cls{TUTA} run on \textbf{Machine~1}, a Dell PowerEdge R730 with two Intel Xeon E5-2640~v4 processors (40 logical cores at 2.40~GHz), 128~GB of DDR4 ECC RAM, and \textbf{no GPU}.
\cls{TUTA} needs a GPU and we ran it on \textbf{Machine~2}, a Dell PowerEdge R770 having two Intel Xeon 6505P processors (48 logical cores), 256~GB of DDR5 RAM, and one NVIDIA RTX PRO 6000 Blackwell Server Edition GPU (96~GB VRAM).
To estimate equivalent cloud costs, we map both machines to the closest Amazon EC2 On-Demand instances available in \texttt{us-east-1}:
\texttt{m5.8xlarge} (32~vCPUs, 128~GiB RAM, \$1.536/hr) for Machine~1 (denoted \textbf{C}) and
\texttt{g7e.8xlarge} (32~vCPUs, 256~GiB RAM, 1$\times$ NVIDIA RTX PRO 6000 Blackwell, \$5.268/hr) for Machine~2 (\textbf{G}).
\printIfExtVersion{Prices correspond to Linux on-demand rates with per-second billing (60-second minimum). Cost estimates assume a single instance running for the full experiment duration.}{}
Tables~\ref{tab:ctc_train_cost} and~\ref{tab:ctc_infer_cost} report runtimes and estimated costs over the full 5-fold cross-validation.
File loading and feature extraction (\emph{Load}) are reported separately from model computation (\emph{Train}/\emph{Infer}), although both
contribute to the total billed time.
\begin{table}
    \centering
    \small
    \caption{Training time and estimated cloud cost (5-fold total). \textbf{C} = \texttt{m5.8xlarge} (\$1.536/hr); \textbf{G} = \texttt{g7e.8xlarge} (\$5.268/hr).}
    \vspace{-1em}
    \setlength{\tabcolsep}{4pt}
    \begin{tabular}{lrrrcr}
        \toprule
        \textbf{Model} & \textbf{Load (s)} & \textbf{Train (s)} & \textbf{Total (h)} & \textbf{Inst.} & \textbf{Cost (\$)} \\
        \midrule
        \cls{RF-Koci}      &  256.7 &     31.6 &  0.08 & C &   0.12 \\
        \cls{RF}           &  252.2 &    120.3 &  0.10 & C &   0.16 \\
        \cls{LightGBM}     &  254.7 &    177.1 &  0.12 & C &   0.18 \\
        \cls{CRF-Linear}   &  252.3 & 14\,658.2 &  4.14 & C &   6.36 \\
        \cls{CRF-RF}       &  250.5 & 15\,848.4 &  4.47 & C &   6.87 \\
        \cls{CRF-LightGBM} &  254.3 & 14\,952.5 &  4.22 & C &   6.49 \\
        \midrule
        \cls{TUTA}         &  178.5 & 92\,376.0 & 25.71 & G & 135.44 \\
        \bottomrule
    \end{tabular}
    \label{tab:ctc_train_cost}
\end{table}
\begin{table}
    \centering
    \small
    \caption{Inference time and estimated cloud cost (5-fold total).
             $^\dagger$CPU-only run of \cls{TUTA} for reference.}
    \vspace{-1em}
    \setlength{\tabcolsep}{4pt}
    \begin{tabular}{lrrrcr}
        \toprule
        \textbf{Model} & \textbf{Load (s)} & \textbf{Infer (s)} & \textbf{Total (h)} & \textbf{Inst.} & \textbf{Cost (\$)} \\
        \midrule
        \cls{RF-Koci}           &  540.0 &     40.0 & 0.16 & C &  0.25 \\
        \cls{RF}                &  555.2 &     49.0 & 0.17 & C &  0.26 \\
        \cls{LightGBM}          &  541.8 &     95.9 & 0.18 & C &  0.27 \\
        \cls{CRF-Linear}        &  540.1 &    268.5 & 0.22 & C &  0.35 \\
        \cls{CRF-RF}            &  550.6 &    300.8 & 0.24 & C &  0.36 \\
        \cls{CRF-LightGBM}      &  542.8 &    288.7 & 0.23 & C &  0.35 \\
        \cls{TUTA}$^\dagger$    &  363.8 & 30\,994.6 & 8.71 & C & 13.38 \\
        \midrule
        \cls{TUTA}              &  203.4 &    679.1 & 0.25 & G &  1.29 \\
        \bottomrule
    \end{tabular}
    \label{tab:ctc_infer_cost}
\end{table}
The results show a clear computational gap between the proposed systems and \cls{TUTA}. Training all six non-DL
systems combined costs~$<$\$21, while \cls{TUTA} alone exceeds \$135 due to extended GPU training.
Inference costs are smaller overall, but CPU-only \cls{TUTA} inference remains much slower and
costlier than GPU inference. All non-DL systems complete inference within 15 minutes at negligible cost.

\begin{figure}
    \centering
    \includegraphics[width=\linewidth]{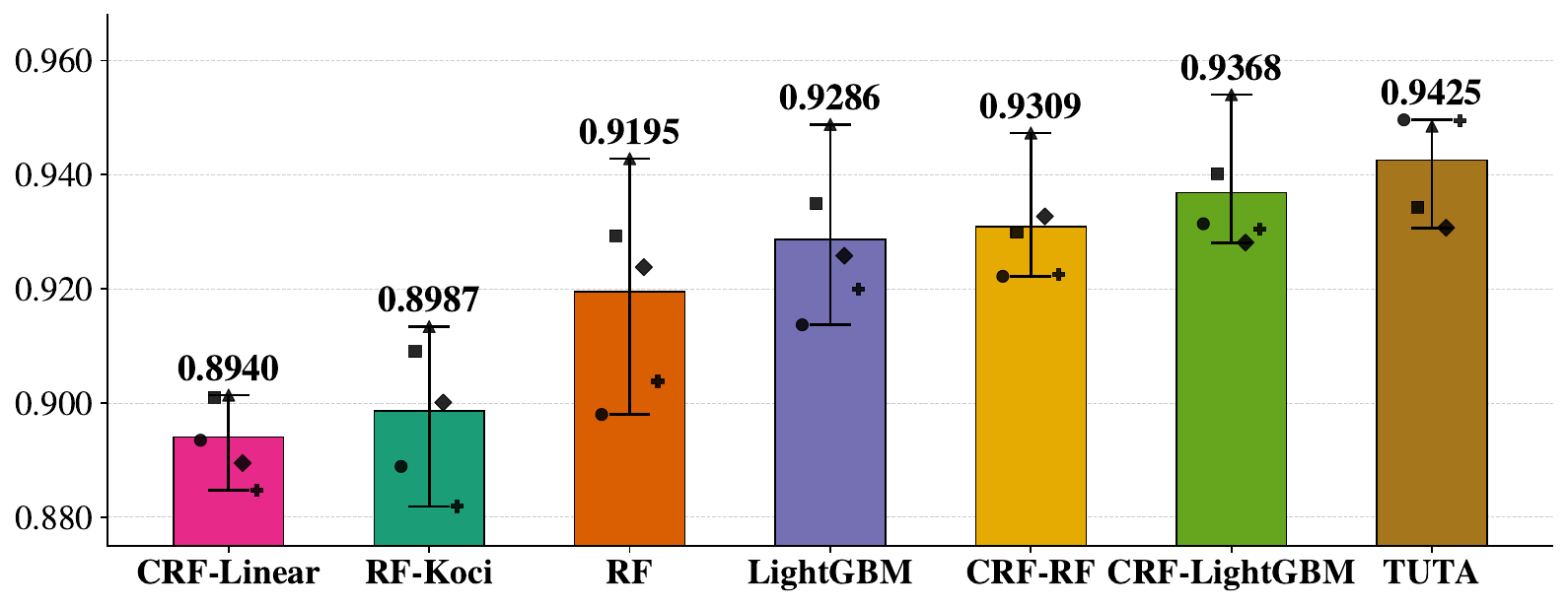}
    \vspace*{-2.5em}
    \caption{Mean File-Macro F$_1$ per model. Error bars span per-fold min/max; individual fold scores are shown as markers.}
    \label{fig:mfm_f1}
\end{figure}

\begin{figure}
    \centering
    \includegraphics[width=\linewidth]{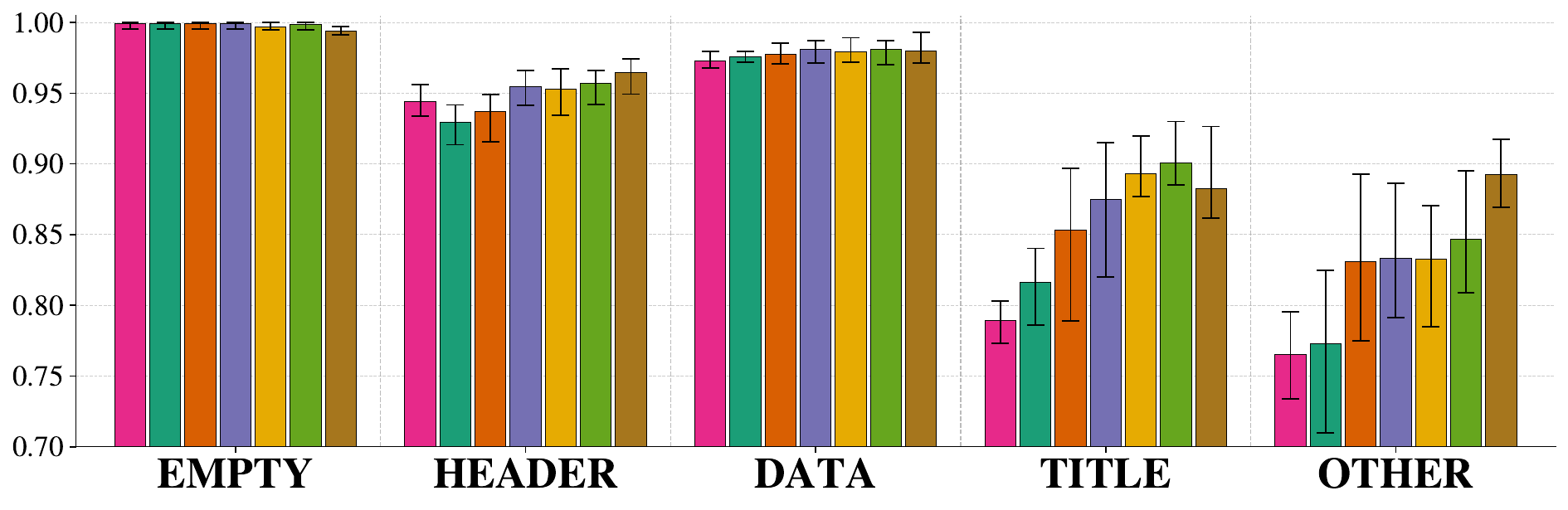}
    \vspace*{-2.5em}
    \caption{Per-class file-macro F$_1$ breakdown; error bars span
    per-fold min/max.}
    \label{fig:per_class_f1}
\end{figure}

\bparagraph{CTC Results}
Figure~\ref{fig:mfm_f1} reports the M-FM-F1 scores of all seven systems averaged over the 5 folds.
Markers indicate per-fold values, and error bars, the min--max range. Figure~\ref{fig:per_class_f1}
shows the corresponding per-class FM-F1$_c$ scores using the same color scheme.

\emph{Effect of spatial awareness, non-linearity, and sequential modeling.}
We assess the contribution of each component comparing successive bars in Figure~\ref{fig:mfm_f1}.
\cls{CRF-Linear} (sequential, spatial, linear)
achieves an M-FM-F1 of 0.894, close to \cls{RF-Koci} (0.899), non-linear but with minimal spatial awareness and
no sequential modeling. Adding full spatial features (\cls{RF-Koci}~$\to$~\cls{RF}) improves performance by $+2.1$ pp (0.899~$\to$~0.920).
Adding sequential modeling through the CRF layer (\cls{RF}~$\to$~\cls{CRF-RF}) yields another $+1.1$ pp, while replacing RF with LightGBM
(\cls{CRF-RF}~$\to$~\cls{CRF-LightGBM}) adds $+0.6$ pp. Overall, \emph{all three components contribute positively and combine effectively}.
\cls{TUTA} reaches %
0.943, only $+0.6$ pp above \cls{CRF-LightGBM} (0.937).

\emph{Per-class analysis.}
The overall gain from \cls{RF-Koci} to \cls{CRF-LightGBM} is $+3.8$ pp in M-FM-F1, but improvements are uneven across classes.
As shown in Figure~\ref{fig:per_class_f1}, \cls{EMPTY} is almost perfectly classified by all systems,
while \cls{DATA} is already saturated (FM-F1~$\approx$~0.976--0.981). %
Gains are larger for harder classes: \cls{HEADER} improves by $+2.8$ pp, \cls{TITLE} by $+8.5$ pp (0.816~$\to$~0.901),
and \cls{OTHER} by $+7.4$ pp (0.773~$\to$~0.847). These classes are harder to distinguish using formatting alone
and benefit from added structural context provided by spatial and sequential modeling.
\cls{TUTA} performs notably better on \cls{OTHER} ($+4.5$ pp),
the most semantically dependent class, but slightly worse on \cls{TITLE} ($-1.9$ pp),
which relies more on structural and formatting cues. \printIfExtVersion{The systems perform similarly on \cls{EMPTY}, \cls{DATA}, and \cls{HEADER}.}{}
These results suggest that \emph{semantic modeling mainly
benefits the most content-dependent class}, while \emph{structural and non-linear modeling already capture most of the remaining signal}.
However, \cls{TUTA}'s benefits come at a substantially higher computational cost.

\subsection{Table Detection}
\label{sec:exp_td}

We compare six \printIfExtVersion{systems for the TD task, including three variants of our deterministic approach
from Sec.~\ref{sec:approach_td} applied on the outputs of different CTC systems,
two region-based spreadsheet partitioning methods, and one LLM-based table detector.
Among these systems, only the LLM-based approach requires task-specific TD training.
Our methods indirectly rely on learning through the CTC component, but the TD stage itself is fully deterministic.
Together with the two region-based baselines, these systems are therefore evaluated directly on each fold without TD-specific training.}{TD systems, as follows.}

\noindent\textbf{\cls{TD(RF-Koci)}} and \textbf{\cls{TD(CRF-LightGBM)}} apply the deterministic \textsc{Table\-RangeExtractor}
(Algorithm~\ref{alg:td_main}, Sec.~\ref{sec:approach_td}) to the grids predicted by \cls{RF-Koci} and
\cls{CRF-LightGBM}, respectively. \printIfExtVersion{Default hyperparameters from Table~\ref{tab:td_hparams} are used.}{}

\noindent\textbf{\cls{TD(Oracle)}} applies \textsc{TableRangeExtractor}
to ground-truth cell-type grids, providing an upper
bound on TD performance under perfect CTC predictions and isolating the TD stage from CTC errors.

\noindent\textbf{\cls{Connected-Component (CC)}} is the region-based baseline from
\cite{vitagliano2021detecting}, following%
~\cite{coletta2012public}.
We use the implementation from the Mondrian repository~\cite{mondrian_github}.  %
As discussed when presented in Sec.~\ref{sec:related_work}, \cls{CC} detects contiguous \emph{regions} rather than semantic tables,
potentially splitting tables or merging them with metadata; it is
evaluated in \emph{region-anchored} mode, yielding an optimistic precision estimate.

\noindent\textbf{\cls{Mondrian}}~\cite{vitagliano2021detecting} is the unsupervised, training-free approach
for spreadsheet region detection and layout template inference presented in Sec.~\ref{sec:related_work}.
Like \cls{CC}, \cls{Mondrian} detects generic regions rather than semantic tables and is therefore evaluated
in \emph{region-anchored} mode. We use the authors' region detection implementation~\cite{mondrian_github},
specifically the \emph{dynamic} variant,
which selects the clustering radius independently for each file, as this setting achieved the best results
in the original work~\cite{mondrian_github}; also, we noted substantial variability in suitable radii across spreadsheets.
\cls{Mondrian} only operates on CSV/TSV files and deliberately ignores formatting information.
\printIfExtVersion{The authors compared
against \cls{TableSense}~\cite{dong2019tablesense} and a genetic-algorithm-based
approach~\cite{koci2019genetic}, but neither code nor trained models remain publicly available.}{}

\begin{table}[t]
  \centering
  \small
  \caption{Inference time and estimated cloud cost for TD (5-fold total).
           \emph{Train.}: training time for \cls{SpreadsheetLLM}; loading and CTC
           inference time for \cls{TD} variants.
           $^\dagger$Assumes perfect CTC at no cost.
           $^\ddagger$Fold 5 timed out after 7 days.}
  \vspace{-1em}
  \setlength{\tabcolsep}{4pt}
  \resizebox{\columnwidth}{!}{\begin{tabular}{lrrrcr}
      \toprule
      \textbf{Model} & \textbf{Train. (s)} & \textbf{Infer. (s)} & \textbf{Total (h)} & \textbf{Inst.} & \textbf{Cost (\$)} \\
      \midrule
      \cls{SpreadsheetLLM} (train) & 33\,096.3 &     ---   & 9.19 & G & 48.43 \\
      \cls{SpreadsheetLLM} (infer) &        --- & 1\,954.8 & 0.54 & G &  2.86 \\
      \midrule
      \cls{TD(RF-Koci)}     &       580.0 &      301.5 &  0.24 & C &  0.38 \\
      \cls{TD(CRF-LightGBM)}&       831.5 &      326.3 &  0.32 & C &  0.49 \\
      \cls{TD(Oracle)}$^\dagger$ & ---   &      355.9 &  0.10 & C &  0.15 \\
      \midrule
      \cls{CC}              &          --- &  2\,723.7 &  0.76 & C &  1.16 \\
      \cls{Mondrian}$^\ddagger$ &      --- & $>$653\,896 & $>$181.6 & C & $>$279.00 \\
      \bottomrule
  \end{tabular}}
  \label{tab:td_cost}
\end{table}

\noindent\textbf{\cls{SpreadsheetLLM (Mistral-7B)}}~\cite{dong2024encoding} is an LLM-based framework for spreadsheet
understanding, evaluated on TD and question answering. Its core component, \emph{SheetCompressor}, converts spreadsheet
grids into compact token sequences by ($i$)~identifying heterogeneous boundary rows and columns;
($ii$)~encoding the cells located on or near these boundaries (within a fixed neighborhood of $k$ rows/columns)
into an inverted-index representation mapping values
to compressed cell addresses. %
Lacking a complete implementation, we built one in Python (available in~\cite{github_repository}), using the partial C\# source code provided in the
supplementary material as a starting point for the \emph{SheetCompressor}.
\printIfExtVersion{The released code omits several filtering functions and requires substantial
adaptation to operate over the variety of spreadsheets we have in \datasetname{}. We reimplement all
documented components while staying as close as possible to the described behavior.}{}
The original work reports %
better results fine-tuning proprietary LLMs.
We choose to fine-tune \verb|mistralai/Mistral-7B-Instruct-v0.2|,
which achieves the best performance among the open models reported by the authors, motivated by reproducibility and cost considerations. We follow the hyperparameters from
\printIfExtVersion{\cite[Appendix~G]{dong2024encoding}:
LoRA adaptation (rank 32, $\alpha=64$, dropout 0.01) on all linear projection layers, AdamW with learning rate $5 \times 10^{-5}$
and cosine scheduling, batch size 5 with 8 gradient accumulation steps (effective batch size 40), and a sequence cutoff of 5\,800 tokens.
Unlike the original setup, which trains for 15 epochs on \cls{WebSheet10K} ($\sim$10k sheets), we use early stopping (patience 3) as validation
loss stabilizes after one to two epochs on our smaller training folds ($\sim$550 sheets). Fine-tuning on our own folds is also necessary
to ensure compatibility with our annotation
protocol.}{\cite[Appendix~G]{dong2024encoding}.}
The model predicts table ranges in Excel notation
(e.g., \verb|A1:B5|). Even after fine-tuning, \printIfExtVersion{predictions
occasionally deviate from the expected format, producing quoted ranges,
malformed coordinates, or isolated cell references. We therefore apply a
lightweight post-processing step based on regular expressions to recover
valid ranges while rejecting ambiguous outputs.
This highlights a practical limitation of generative approaches for structured prediction,
where additional recovery logic remains necessary despite task-specific fine-tuning.}{some predictions are malformed, requiring post-processing
to recover valid results and reject those which cannot be interpreted: a practical limitation of generative approaches for structured prediction.}

\bparagraph{TD Metrics}
A table is represented as a rectangular bounding box defined by its row and column extents
in the cell grid. For a predicted box $\hat{t}$ and a ground-truth box $t$,
we measure overlap with the standard %
$
  \text{IoU}(\hat{t}, t) = \frac{|\hat{t} \cap t|}{|\hat{t} \cup t|}
$
 where areas are counted in cells.

\emph{Precision curve.}
We report precision as a function of the IoU threshold $\tau \in [0, 1]$.
At threshold $\tau$, a matched pair $(\hat{t}, t)$ is counted as a true
positive if $\text{IoU}(\hat{t}, t) \geq \tau$, and as a false positive
otherwise. Precision at $\tau$ is thus
\(
  \text{Precision}(\tau) = \frac{|\{(\hat{t},t) : \text{IoU}(\hat{t},t) \geq
\tau\}|}{N_{\text{pred}}}
\).
We report both \textbf{micro} precision (pooling all predictions across
sheets) and \textbf{macro} precision (averaging per-sheet precision
\printIfExtVersion{curves), sweeping $\tau$ continuously over $[0, 1]$.}{curves.}

\emph{Matching.} Predicted and ground-truth tables are matched greedily in
decreasing order of IoU: each table on either side is matched at most once (tables
cannot overlap), and pairs with IoU $= 0$ are not matched. Unmatched predicted
tables count as false positives; unmatched ground-truth tables count as false negatives.
Systems that predict tables directly use $N_{\text{pred}}$ as the total number of
predictions (\textit{table-anchored} mode). For region-based systems (\cls{CC} and \cls{Mondrian}),
evaluated in \textit{region-anchored} mode, $N_{\text{pred}}$ is
capped at $\min(N_{\text{pred}}, N_{\text{gt}})$ to avoid penalizing
over-segmentation that is intrinsic to the region detection paradigm. This favors
region-based methods by discarding excess predictions, yielding an optimistic precision estimate.
In both modes, since matching is one-to-one,
every predicted table is either matched or unmatched, yielding
$\text{Recall}(\tau) = \text{Precision}(\tau) \times N_{\text{pred}} / N_{\text{gt}}$
for the micro curve: recall is a constant rescaling of precision, thus we do not report it.

\bparagraph{TD Computational Efficiency}
Experimental infrastructure and cost estimation follow Sec.~\ref{sec:exp_ctc},
using the same instances
\printIfExtVersion{mappings: \textbf{C} (\texttt{m5.8xlarge}, \$1.536/hr) and
\textbf{G} (\texttt{g7e.8xlarge}, \$5.268/hr)}{\textbf{C} and \textbf{G}}. \cls{SpreadsheetLLM} requires a GPU
and runs on Machine~2 (\textbf{G}); others run on Machine~1 (\textbf{C}).
Table~\ref{tab:td_cost} reports runtimes and estimated costs over the full
5-fold cross-validation. \printIfExtVersion{Due to the heterogeneous nature of the compared systems,
columns have different interpretations.}{} For \cls{SpreadsheetLLM}, \emph{Train.} and \emph{Infer.}
correspond to fine-tuning and inference. For \cls{TD} variants, \emph{Train.}
reports \emph{loading and CTC inference time} of the underlying CTC model (in Table~\ref{tab:ctc_infer_cost}),
while \emph{Infer.} measures the deterministic TD stage alone. \cls{TD(Oracle)} has zero training cost as
perfect CTC predictions are assumed (unrealistic upper-bound). For \cls{CC} and \cls{Mondrian}, only inference time is reported,
including file loading.
For \cls{Mondrian}, we interrupted fold~5 after 7 days (604\,800~s)
spent processing a spreadsheet with over 1M cells.
The reported values combine those for folds~1 to 4 (49\,095.9~s) with
the full timeout duration; they are lower bounds.

The computational cost of TD is dominated by two
outliers. \cls{Mondrian} is by far the most expensive system due to
the 7-day timeout in fold~5, yielding a lower-bound cost of \$279. Excluding this, total cost is \$53.47, of which \$51.29 comes from \cls{SpreadsheetLLM} alone due to GPU fine-tuning.
By contrast, our \cls{TD} variants are inexpensive: \cls{TD(RF-Koci)}/\cls{TD(CRF-LightGBM)} cost \$0.38/\$0.49 (CTC inference included), and \cls{CC} costs \$1.16.
Accounting for both CTC training and inference, the full \cls{TD(CRF-LightGBM)}
pipeline amounts to \$6.98 in total. Overall, the deterministic TD pipeline adds little overhead on top of the underlying CTC system and remains
substantially cheaper than both GPU-based and unsupervised alternatives.

\begin{figure}[t]
  \centering
  \includegraphics[width=0.9\linewidth]{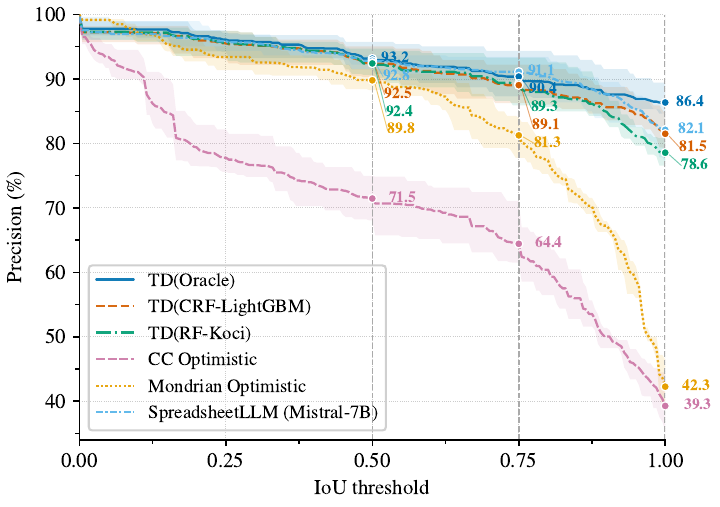}
  \vspace*{-1.7em}
  \caption{Macro-averaged (per-sheet) TD precision. Solid lines show fold
  means; shaded areas indicate min--max.}
  \label{fig:macro_precision}
\end{figure}

\begin{figure}
  \centering
  \includegraphics[width=0.9\linewidth]{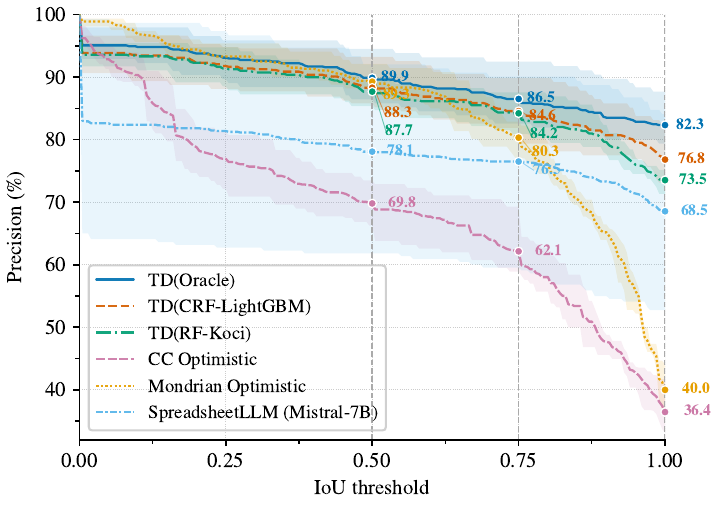}
  \vspace*{-1.7em}
  \caption{Micro-averaged TD precision. Solid lines show fold means;
  shaded areas indicate min--max.}
  \label{fig:micro_precision}
\end{figure}

\bparagraph{TD Results}
Figures~\ref{fig:macro_precision} and~\ref{fig:micro_precision} report macro and micro precision
as a function of $\tau$ for all systems. All methods consistently achieve higher macro
than micro precision, indicating over-prediction of tables in larger/more complex spreadsheets,
where micro metrics are more~sensitive.%

\emph{Effect of CTC quality on table detection.}
The \cls{TD} variants directly reflect the quality of their underlying CTC systems. Replacing
\cls{RF-Koci} with \cls{CRF-LightGBM} improves precision at IoU~$=$~1 by $+2.9$~pp on macro (78.6~$\to$~81.5\%)
and $+3.3$~pp on micro (73.5~$\to$~76.8\%). As shown in Figures~\ref{fig:macro_precision} and~\ref{fig:micro_precision},
these gains mainly appear at high IoU thresholds, indicating more accurate boundary recovery rather than better coarse
localization. This is consistent with the CTC results of Sec.~\ref{sec:exp_ctc}: \cls{DATA} and \cls{EMPTY}
cells are already well classified by all systems, while remaining errors mostly affect boundary-related classes
such as \cls{TITLE}, \cls{HEADER}, and \cls{OTHER}.
Using perfect CTC predictions, \cls{TD(Oracle)} reaches 86.4\% macro and 82.3\% micro precision at IoU~$=$~1.
Relative to \cls{TD(CRF-LightGBM)}, this leaves a margin of $+4.9$~pp and $+5.5$~pp respectively,
suggesting that CTC quality gains would also translate into better table detection.

\emph{Region-based methods.}
Despite the optimistic evaluation protocol for \cls{CC} and \cls{Mondrian},
both region-based approaches remain substantially below \cls{TD} variants at high IoU thresholds.
At IoU~$=$~1, \cls{Mondrian} reaches 42.3\% macro and 40.0\% micro precision, trailing \cls{TD(CRF-LightGBM)}
by $-39.2$~pp and $-36.8$~pp. However, at lower IoU thresholds, it performs comparably
to several other systems, indicating that its predicted regions often overlap tables but fail to recover
precise boundaries. These results suggest that \emph{generic region detection is insufficient for precise spreadsheet
table extraction}.

\emph{Comparison with \cls{SpreadsheetLLM}.}
\cls{SpreadsheetLLM} is close to \cls{TD(CRF-LightGBM)} in macro precision
($+0.6$~pp at IoU~$=$~1\printIfExtVersion{, 81.5 $\to$ 82.1\%}{}), but substantially worse on micro precision
($-8.3$~pp, 76.8 $\to$ 68.5\%). This larger macro/micro gap suggests \emph{stronger over-prediction
on complex spreadsheets}. We also observe \emph{higher variance across folds}, indicating greater
sensitivity to the training distribution. In addition,  model outputs are occasionally inconsistent with the requested output format and require post-processing. %
Overall, \cls{TD(CRF-LightGBM)} achieves competitive results while remaining substantially
cheaper than \cls{SpreadsheetLLM}. The oracle results
further indicate that additional gains remain possible through improved CTC predictions.

\section{Conclusion}
\label{sec:conclusion}

High-quality, cost-efficient and reproducible methods for extracting
information from spreadsheets are still needed. We provide a
comprehensive annotated dataset and propose a new CTC-TD pipeline
combining strong performance, low resource needs, and full TD
interpretability. Our CTC method, combining spatial, sequential,
and non-linear features, and our deterministic, learning-free TD method, match or
outperform state-of-the-art DL competitors (\cls{TUTA} for CTC,
\cls{SpreadsheetLLM} for TD) at a fraction of their cost and
without requiring a GPU. With no black-box effects in TD,
our corpus and pipeline advance the state of the art
in spreadsheet understanding.

\begin{acks}
This work was funded in part by the French government under management of
Agence Nationale de la Recherche (ANR) as part of the ``France~2030'' program,
references ANR-23-IACL-0008 (PR[AI]RIE-PSAI) and ANR-23-IACL-0005 (Hi! PARIS Cluster 2030).
\end{acks}

\versioning{\clearpage}{}

\section*{GenAI Usage Disclosure}

We used an LLM tool
(\href{https://claude.com/product/claude-code}{\emph{Claude Code}},
running Claude Sonnet 4.6) during the development of this work,
on two occasions.

First, we used it to assist in the reproduction of the \cls{SpreadsheetLLM} system described
in~\cite{dong2024encoding}, for which no full implementation code was provided.
In particular, we leveraged it to reconstruct the \emph{SheetCompressor}
component from the C\# code fragments provided in their supplementary material, to translate them into Python, and
integrate them with the rest of the system's code. We also used Claude to
re-implement the overall pipeline of \cite{dong2024encoding} based on the
methodological details described in the paper and its appendices. All generated or assisted code was manually reviewed
and corrected when necessary. We verified correctness against the paper descriptions and ensured that no modifications
altered the intended behavior of the original algorithms. 

In addition, we used the LLM to clean our own codebase and improve its organization, as well as to refine CLI
descriptions to enhance usability for reproducing the experiments. All such changes were reviewed to ensure
behavioral equivalence and consistency with the experimental setup described in the paper.

No generative AI system was used to write any part of the current paper.

\bibliographystyle{ACM-Reference-Format}
\bibliography{main}

\end{document}